\documentclass[a4paper,11pt]{article}
\pdfoutput=1 

\usepackage{jheppub} 

\usepackage[T1]{fontenc} 

\usepackage{physics}
\usepackage{subcaption}

\title{\boldmath The Holographic Entropy Cone in AdS-Vaidya Spacetimes}

\author{Reginald J. Caginalp}

\affiliation{Department of Physics,\\University of California,\\Berkeley, CA 94720, USA}

\emailAdd{caginalp@berkeley.edu}

\abstract{We examine the five-region holographic entropy cone inequalities for the special case of the AdS$_3$-Vaidya metric for a variety of boundary configurations. This is done by numerically solving the geodesic equation in the bulk for various boundary configurations. In all the cases we examine, we find that all the inequalities are satisfied when the bulk satisfies the null energy condition, while the inequalities are all violated when the bulk spacetime violates the null energy condition. A proof of the five-region holographic entropy cone inequalities for the dynamical bulk case remains an open problem--our results provide evidence that these inequalities hold for dynamical bulk spacetimes.}

\begin{document} 
\maketitle
\flushbottom

\section{Introduction}

Recent work has unveiled deep connections between gravity and entanglement. The AdS/CFT correspondence \cite{Maldacena, Gubser, Witten}  states that any theory of quantum gravity in $(d+1)$-dimensional anti-de Sitter space (AdS$_{d+1}$) is equivalent to a conformal field theory (CFT) in $d$ dimensions. The Ryu-Takayanagi (RT) \cite{RT} formula and its covariant generalization, the Hubeny-Rangamani-Takayanagi (HRT) formula \cite{HRT}, posit that the entanglement entropies of holographic CFTs are given in terms of minimal or extremal areas. These have been derived from the basic AdS/CFT dictionary \cite{RT_deriv, HRT_deriv}. In general, entanglement entropies of quantum field theories are difficult to compute. It is of great interest to try to determine which types of states are dual to semi-classical AdS bulks. The fact that holographic entanglement entropies are given by minimal areas should therefore enable us to constrain the entanglement structure of holographic states.  

It is simple to show that if we have three spatial CFT regions, $A, B, C$, then the RT formula implies strong subadditivity \cite{RT_SSA,MMI_SSA2}: 
$$S(ABC)+S(B) \leq S(AB) + S(BC).$$
The above inequality is, of course, true for all quantum states \cite{Lieb1,Lieb2}, though the general proof is technically complicated.
In addition, holographic entropies obeying the RT formula obey the constraint of monogamy of mutual information \cite{RT_MMI,MMI_SSA2}: 
$$I(A:BC) \geq I(A:B) + I(A:C).$$
Unlike strong subadditivity, this constraint is not obeyed by all quantum systems. In addition, recent work \cite{Cone} has shown that holographic entanglement entropies for $n$ regions obey a set of inequalities known as the \textit{holographic entropy cone}, assuming the RT formula holds. Recently, the exact holographic entropy cone for five regions has been obtained \cite{Cuenca}. However, it is not known in general if these inequalities are valid for the covariant HRT formula. 

Using the maximin formalism of Wall \cite{Wall}, it is possible to show that (assuming the null-energy condition holds in the bulk), strong subadditivity and monogamy of mutual information hold for the HRT formula. However, the validity of the inequalities for the entropy cone for the HRT formula for five or more regions remains unknown. Indeed, \cite{Rota} showed that the set of five-region inequalities provable with the maximin formalism is less strong than the entropy cone inequalities. 

Understanding the validity of the entropy cone inequalities in the dynamical, HRT case is thus an important step towards understanding the structure of holographic states. In this paper, we will numerically calculate the entanglement entropies for an AdS$_3$-Vaidya spacetime, and examine the validity of the five-region entropy cone inequalities, using the HRT formula. This is a very simple setting to test these inequalities, since the HRT surfaces will be geodesics (not higher-dimensional surfaces), and the AdS$_3$-Vaidya solution is a very simple dynamical spacetime. 

We find that the inequalities are all valid, in the cases we examined, as long as the bulk obeys the null energy condition. If the bulk violates the null energy condition, then all the inequalities are violated. This is analogous to the situation for strong subadditivity, which requires the NEC to hold in the bulk.  We believe that this provides strong evidence for the validity of the five-region inequalities when the bulk is dynamical. Moreover, the shape of the curves resemble those of the strong subadditivity curves.  This may hint that there is a reformulation of the HRT prescription for which both strong subadditivity and the five-region inequalities are valid. Indeed, this has already been done in certain limits for the positive-energy spacetime we considered here \cite{BaoMezei}. 

Understanding the validity of these inequalities in general, as well as further study of the entanglement of holographic states, will be very important in furthering our understanding of quantum gravity.  

Our results build on previous work on the validity of the five-region inequalities for dynamical bulks. \cite{Flory} numerically verified the inequalities for a holographic model of two 1+1 dimensional heat baths joined at $t=0$. It has also been shown that they are valid for large, late-time CFT regions in collapsing black hole spacetimes \cite{BaoMezei}. Our work is closely analogous to \cite{Headrick,Allais}, which numerically studied the validity of strong subadditivity and monogamy of mutual information for AdS$_3$-Vaidya spacetimes.  
 
\section{Setup}
Consider a holographic CFT with a Cauchy slice $\Sigma$ of a static bulk, at a moment of time reflection symmetry. Let $A$ be a boundary subregion.  The Ryu-Takayanagi formula posits that
$$S(A) = \frac{ \min_{m} \text{Area}}{4 G_N},$$
where $m$ is a codimension-2 surface in the bulk (with $\partial m = \partial A$) homologous to $A$. That is, there is a bulk region $\chi$ such that $\partial \chi = A \cup m$. The Hubeny-Rangamani-Takayanagi formula is the covariant generalization of this equation. If $A$ is some spacelike CFT subregion, then the HRT formula says that 
$$S(A)=  \frac{ \min_{\text{extremal } m} \text{Area}}{4 G_N},$$
where extremal $m$ means that $m$ is a co-dimension 2 spacelike surface that extremizes the area and has $\partial m = \partial A$ and is homologous to $A$.   

We will consider the planar AdS$_3$-Vaidya spacetime, with metric 
$$ds^2=-(r^2-m(v)) dv^2+ 2 dr dv +r^2 dx^2.$$
We wish to see when the null-energy condition (i.e., $T_{\mu \nu} k^\mu k^\nu \geq 0$ for all lightlike $k$) is satisfied for this metric. The Einstein tensor of this metric is given by 
$$G_{v r} = G_{r v} =1, G_{x x} =r^2, G_{v v}=-r^2+m(v)+\frac{1}{2r} \frac{dm}{dv},$$
with all other components vanishing. We need to find the null vectors $k^ \mu$ in the $r$-$v$ plane. Suppose $k^\mu = (C,D,0)$ so that we require 
$$-f(r,v)C^2 +2CD =0,$$
where we have introduced the function $f(r,v) \equiv r^2-m(v)$. There are two solutions to this:
$$ C =0 \text{  or  } D=\frac{f}{2}C,$$
so there are two such linearly-independent null vectors 
$$ n^\mu = (0,1,0), \ell^\mu = (1, \frac{f}{2},0).$$
Einstein's equations are 
$$ G_{\mu \nu } + \Lambda g_{\mu \nu} = 8 \pi G_N T_{\mu \nu}.$$
These imply 
$$T_{\mu \nu} n^\mu n^\nu=G_{\mu \nu} n^\mu n^\nu=G_{rr}=0,$$
$$T_{\mu \nu} \ell^\mu \ell^\nu=G_{\mu \nu} \ell^\mu \ell^\nu=G_{vv}+f G_{v r} =-f+f+\frac{1}{2r} \frac{dm}{dv}=\frac{1}{2r} \frac{dm}{dv}.$$
Therefore, we see that the null energy condition is satisfied if and only if $dm/dv$ is positive. 

We can re-write this metric in more standard coordinates, $t$ and $r$, with 
$$v= t+ g(r), g'(r) = \frac{1}{f}$$
$$\implies dv = dt+ g'(r) dr = dt +\frac{dr}{f},$$
which means 
$$ds^2 = - f (dt^2+\frac{dr^2}{f^2} + 2 \frac{dt dr}{f}) +2 dr dt +2 \frac{dr^2}{f} +r^2 dx^2= -f dt^2 + \frac{dr^2}{f} +r^2 dx^2.$$
We will consider a thin-shell limit, 
$$m(v)=\pm m \Theta(v),$$
which represents a shell of infalling null matter. The plus sign satisfies the null energy condition and corresponds to positive energy matter--inside the shell, the metric is pure AdS$_3$, outside the shell, the metric corresponds to a black hole, i.e., the BTZ metric. The minus sign violates the null energy condition; this choice represents a shell of negative-energy null matter so that outside the matter, the metric is pure AdS, while inside it is BTZ. We will consider both cases, starting with the positive-energy metric. For simplicity, we will set $m=1.$ 

For AdS, we have $f=r^2$ so that 
$$g'= -\frac{1}{r} \implies v= t-\frac{1}{r}.$$
Meanwhile for BTZ, we have $g=r^2-1$, which gives
$$g' = -\frac{1}{r^2-1} \implies g = - \tanh^{-1} \frac{1}{r} \implies v = t- \tanh^{-1} \frac{1}{r} .$$

Our discussion of geodesic kinematics largely follows that of \cite{Headrick}. 

\section{Positive Energy Vaidya Metric}

We wish to obtain enanglement entropies in the CFT dual to the Vaidya metric. The HRT prescription tells us that we need to calculate the areas of the extremal codimension 2 surfaces that are anchored at the boundary of the CFT subregion. In our case, this corresponds to spacelike geodesics. 
The geodesic equation for $v$ is given by 
$$\ddot{v}+\frac{1}{2} \partial_r f \dot{v}^2-r \dot{x}^2=0,$$
where an overdot denotes a derivative with respect to the affine parameter $\tau$. In addition, the tangent vector $d x^{\mu} /d \tau $ has unit norm, so that 
$$-f(r,v) \dot{v}^2 +2 \dot{v} \dot{r} + r^2 \dot{x}^2=1.$$
Because the metric has no explicit $x$-dependence, the quantity $p_x \equiv g_{xx} \dot{x} = r^2 \dot{x}$ is conserved along the geodesic. When we express the metric in terms of $t$ and $r$ (instead of $v$ and $r$), we found that the metric takes the form 
$$ds^2 = -h(r) dt^2 + \frac{1}{h(r)} dr^2 +r^2 dx^2,$$
where $h(r)=r^2$ in AdS, and $h(r)=r^2-m$ in BTZ. For simplicity, we will set $m=1$ throughout. Other than at the shell, there is no explicit dependence in the metric on $t$, so there is a quantity that is conserved except at the shell. It is given by
$$E \equiv g_{tt} \dot{t} = h(r) \dot{t}.$$
In these coordinates, the normalization condition for the vector $d x^ \mu /d \tau$ becomes
$$-h(r) \dot{t}^2 + \frac{ \dot{r}^2} {h(r)}+r^2 \dot{x}^2 =1 $$
$$\dot{r}^2=h+E^2-\frac{h p_x^2}{r^2} $$
Alternatively, we can write this in terms of $r'$, where the prime denotes a derivative with respect to $x$. Then we obtain
$$r'^2 = \frac{r^4 h}{p_x^2} + \frac{ r^4 E^2}{p_x^2}-h r^2.$$
We will first solve these equations for constant time intervals, and then solve them for general covariant regions. For both of these, we will verify strong subadditivity and the 5-body entropy cone inequalities. 

\subsection{Constant Time Intervals}

We begin by considering a region $A$ on the boundary that is constant in time, and is a single interval in $x$. That is, 
$$A = \{ (x,t) \in \text{CFT} | x \in [0, \ell_x], t=const=t_b \}.$$
To calculate the entanglement entropy $S(A)$ of this region in the CFT, the HRT prescription tells us that we need to find the extremal boundary-anchored curve (i.e., spacelike geodesic) $\chi_A$ such that $\partial \chi_A = \partial A$ and $\chi_A$ is homologous to $A$.
\begin{figure}
\centering
 \includegraphics[width=0.35\textwidth]{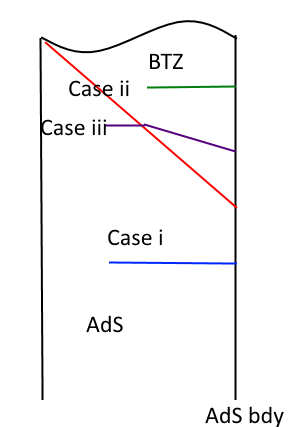}

 \caption{A Penrose diagram of the Vaidya spacetime (the red line represents the shell of null matter), showing the three cases for the spacelike geodesics. (i) Entirely in the AdS bulk, (ii) entirely in the BTZ bulk, and (iii) partially in the BTZ bulk, and partially in the AdS bulk.}\label{fig:Pen_const}
\end{figure}

There are three cases (i) the geodesic is entirely in the AdS bulk, (ii) the geodesic is entirely in the BTZ bulk, or (iii) the geodesic is in both the BTZ bulk and the AdS bulk. See Figure~\ref{fig:Pen_const}. We consider each of these cases in turn. 

\subsubsection{Geodesics Entirely in the AdS Bulk}

We are considering a constant time geodesic in the AdS bulk, so $E=0$. Therefore, we have 
$$\frac{dr}{d \tau} = \sqrt{r^2-p_x^2},$$ 
which has solution
$$r(\tau) = \frac{1}{2}(p_x^2 e^{-\tau} + e^\tau).$$
Now, $x$ obeys the equation 
$$\dot{x} = \frac{p_x}{r^2},$$
which has solution
$$x(\tau) = Const - \frac{2p_x}{p_x^2+e^{2 \tau}}.$$
This means that 
$$\ell_x = x(\tau = \infty ) - x( \tau = - \infty) = \frac{2}{p_x}.$$
We have normalized our affine parameter so that $\tau$ measures the length of the curve. For $\tau$ approaching $ \pm \infty,$ $r$ approaches $\infty$. For large $R$, there are two roots of $\tau$., one large and positive, the other large and negative. They are 
$$\tau_{b-} = - \log ( 2 R )  + 2 \log( p_x )$$
$$\tau_{b+} =  \log ( 2 R ) ,$$
so the total length of the curve is 
$$L = \tau_{b+} - \tau_{b-} = 2 \log (2R) - 2 \log (p_x).$$
To get to the boundary, of course, we need to send $R \rightarrow \infty$, and the length diverges. Thus, to obtain a regularized, finite length, we need to subtract the UV-divergent term. Thus, we obtain:
$$ L_{reg} = 2 \log \frac{\ell_x} {2}.$$
This is a concave function, so it satisfies strong subadditivity. 

\subsubsection{Geodesics Entirely in the BTZ bulk}

In this case we have 
$$ \dot{r} = \sqrt{r^2-1 +E^2 -p_x^2+\frac{p_x^2}{r^2}}.$$
There are two solutions to this equation: 
$$r_1(\tau) = \frac{1}{2} \sqrt{-2 E^2 +2p_x^2+2+e^{2 \tau} -E^4 e^{ -2 \tau}+2E^2(p_x^2+1)e^{ -2 \tau}-(p^2-1)^2 e^{ -2 \tau}},$$
$$r_2(\tau) = \frac{1}{2} \sqrt{(-E^2+2E^2p_x^2+2E^2-p_x^4+2p_x^2-1)e^{2 \tau}-e^{-2 \tau} -2E^2+2p_x^2+2}.$$
With a few lines of algebra, we can cast these in the following form: 
$$r_1(\tau)^2 = \frac{1}{4} (e^\tau + B_+ e^{-\tau})(e^\tau + B_- e^{-\tau}),$$
$$r_2(\tau)^2 = -\frac{1}{4} (B_+e^\tau - e^{-\tau})(B_-e^\tau -e^{-\tau}),$$
where we have defined the quantities
$$B_{\pm} = (p_x \pm 1)^2 -E^2.$$
We are, of course, looking for geodesics that are boundary anchored. As $\tau$ goes to minus infinity, $r_2^2$ goes to $-1/4$. Thus, the solution $r_2$ can never describe the geodesics we are interested in. Therefore, we restrict our attention to the solution $r_1$.  We first obtain expressions for $x$ and $t$. We find 
$$t(\tau) =const+ \frac{1} {2} \log \left ( \frac{A_-+e^{2 \tau}}{A_+ + e^{2 \tau}} \right),$$
$$x(\tau) = const - \frac{1} {2} \log \left ( \frac{B_-+e^{2 \tau}}{B_+ + e^{2 \tau}} \right),$$
where we have defined
$$A_\pm = p_x^2 -(1 \pm E)^2.$$
An analysis exactly the same as above gives us that the regularized length is (subtracting off the UV-divergent term $2 \log (2R)$)
$$L_{reg} = -\frac{1}{2} \log (B_+ B_-).$$
Meanwhile, 
$$\Delta t  = \frac{1}{2} \log(A_-/A_+), \text{		} \ell_x = -\frac{1}{2} \log ( B_- /B_+).$$
In particular, if we have a constant-time interval, then $E=0$ and 
$$L_{reg} = - \frac{1}{2} \log (p_x^2-1)^2 , \ell_x= \frac{1}{2}\log \left ( \frac{p_x+1}{p_x-1} \right ).$$

Finally, we turn to geodesics that are partially in AdS and partially in BTZ.

\subsection{Geodesics in both AdS and BTZ}
Because the shell is located at $v=0$, if the interval is at boundary time $t_b<0$, it will be in pure AdS. If $t_b>0,$ then we calculate $v( \tau) = t(\tau) - \tanh^{-1} (1/r(\tau) )$ for the pure BTZ geodesic, and see if it dips below $0$. If it does, then it will have a component that is in the AdS bulk. If not, it will be contained entirely in the BTZ bulk. 

We begin by considering what happens at the junction of AdS and BTZ. Because we do not want a delta function in $v''$, then $v'$ needs to be continuous. This means:
$$v'= 2(r_A' -r_B').$$
In AdS space, we have that  
$$v'=t'+\frac{r_A'}{r_c^2}=\frac{E_A}{p_x}+\frac{r_A'}{r_c^2},$$
where $r_c$ is the value of $r$ when the geodesic crosses the shell, and $E_A$ is the value of $E$ in the AdS region. We combine these two to get
$$r_B' = -\frac{E_A}{2p_x} +\left ( 1- \frac{1}{2 r_c^2} \right ) r_A' $$
From the BTZ side we know that 
$$ v'=t'+\frac{r_B'}{r_c^2-1}= \frac{r_c^2}{r_c^2-1} \frac{E_B}{p_x}+\frac{r_B'}{r_c^2} =\frac{r_c^2}{r_c^2-1} \frac{E_B}{p_x}+\frac{r_B'}{r_c^2-1}.$$

This gives us:
$$(r_c^2-1) \frac{E_A}{p_x}+r_A' -\frac{r_A'}{r_c^2}=r_c^2 \frac{E_B}{p_x}+r_B'$$
$$(r_c^2-\frac{1}{2}) \frac{E_A}{p_x} -\frac{r_A'}{2 r_c^2}=r_c^2 \frac{E_B}{p_x} $$
$$E_B = (1-\frac{1}{2 r_c^2} ) E_A-\frac{p_x r_A'}{2 r_c^4}.$$
In addition, we know that 
$$r_A'= \sqrt{\frac{r_c^6}{p_x^2}+\frac{r_c^4 E_A^2}{p_x^2}-r_c^4}.$$
The value of the affine paramter in the BTZ geodesic when $r=r_c$ is given by
$$\alpha_c \equiv \exp(2 \tau_c) = \frac{1}{2} \left [ -(B_+ +B_-) + 4r_c^2+\sqrt{-4B_+ B_-+ (B_+ +B_--4r_c^2)^2} \right ].$$
There are two BTZ components. For second component, from $\tau_c$ to $\tau = \infty$, we have 
$$\Delta x_B= x(\tau = \infty) - x(\tau_c)= -\frac{1} {2} \log \left ( \frac{B_-+\alpha_c}{B_+ + \alpha_c} \right),$$
$$\Delta t_B = \frac{1} {2} \log \left ( \frac{A_-+\alpha_c}{A_+ + \alpha_c} \right).$$
At the shell, we have that $t_c=\tanh^{-1} (1/r_c)$, so the boundary time is 
$$t_b= \tanh^{-1} (1/r_c) + \Delta t_B.$$

Meanwhile, the length of the curve is 
$$L_B = \log 2R -\frac{1}{2} \log \alpha_c.$$
By symmetry, the portion of the curve in AdS has $E=0$, and the two BTZ components have the same $\Delta x$ and the same length. The curve in AdS obeys 
$$r(\tau) = \frac{1}{2}(p_x^2 e^{-\tau} + e^\tau)$$
This satisfies $r=r_c$ at two values of the affine parameter 
$$\tau_\pm = \log(r_c \pm \sqrt{r_c^2-p_x^2}).$$
\begin{figure}
\centering
 \includegraphics[width=0.7\textwidth]{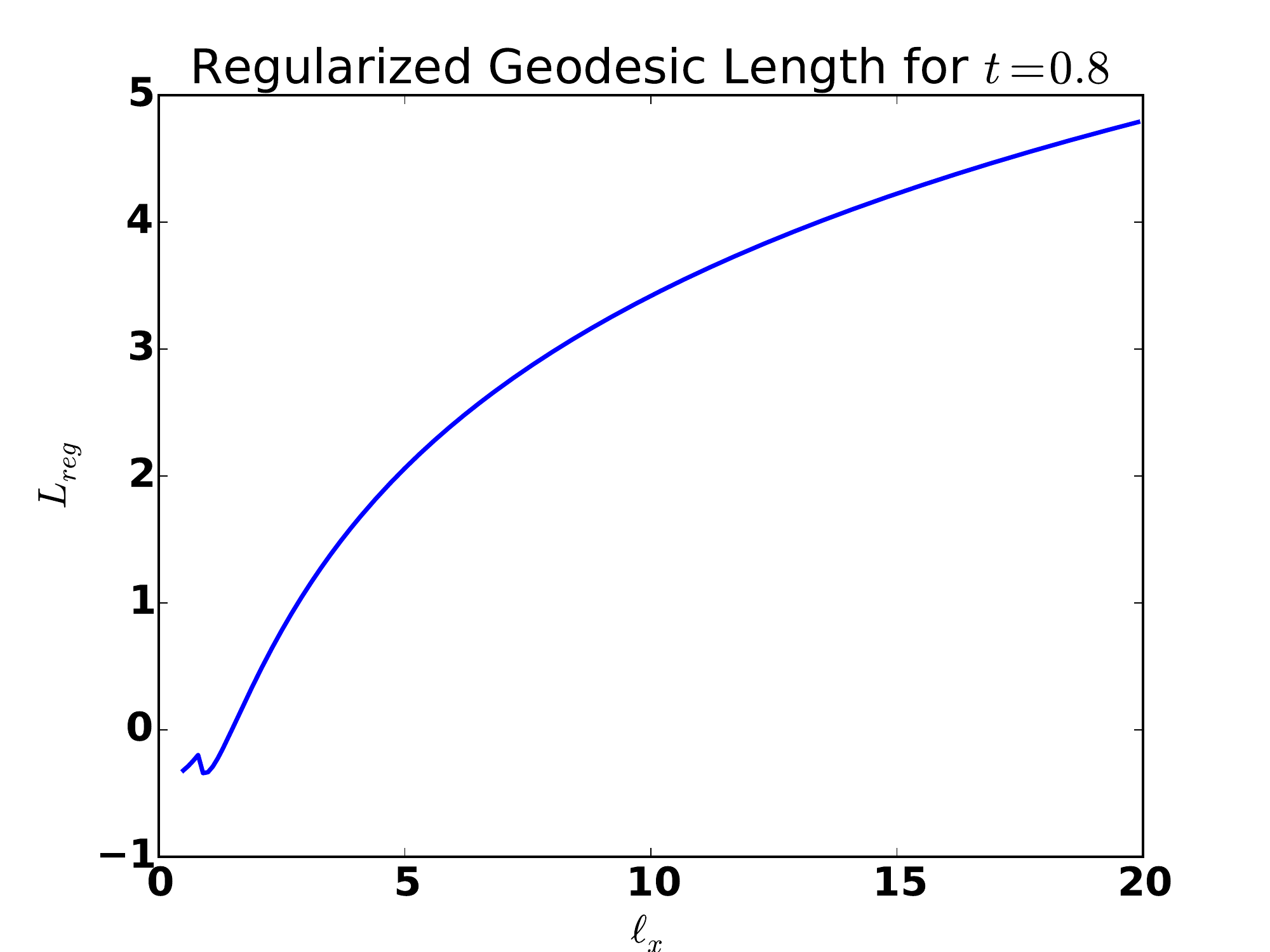}

 \caption{The regularized geodesic length $L_{reg}$ as a function of the boundary interval length, $\ell_x$ at boundary time $t_b=0.8$.}\label{fig:geo_len}
\end{figure}
Using our expression for $x(\tau),$ we find that 
$$\Delta x_A = x(\tau_+ ) - x (\tau_-) =\frac{2}{r_c p_x} \sqrt{r_c^2 -p_x^2}$$
and that 
$$L_A = \tau_+ - \tau_- = \log \left( \frac{r_c+\sqrt{r_c^2 -p_x^2}}{r_c-\sqrt{r_c^2 -p_x^2}}\right).$$
We know that $L=L_A+2L_B$ and $\ell_x = 2 \Delta x_B +\Delta x_A$. We then find 
$$\ell_x  = \frac{2}{r_c p_x} \sqrt{r_c^2 -p_x^2} + \frac{1}{2} \log \left ( \frac{ B_++\alpha_c}{ B_-+\alpha_c}\right ),$$
$$L_{reg} =\log \left( \frac{r_c+\sqrt{r_c^2 -p_x^2}}{r_c-\sqrt{r_c^2 -p_x^2}}\right) -\log \alpha_c .$$

For given values of $t_b$ and $\ell_x$, we can solve numerically to find the corresponding values of $r_c$ and $p_x$. We do this numerically for $t_b=0.8$, and show the result in Figure~\ref{fig:geo_len}. 

\begin{figure}

\centering
\begin{subfigure}{0.45\textwidth}
 \includegraphics[width=\textwidth]{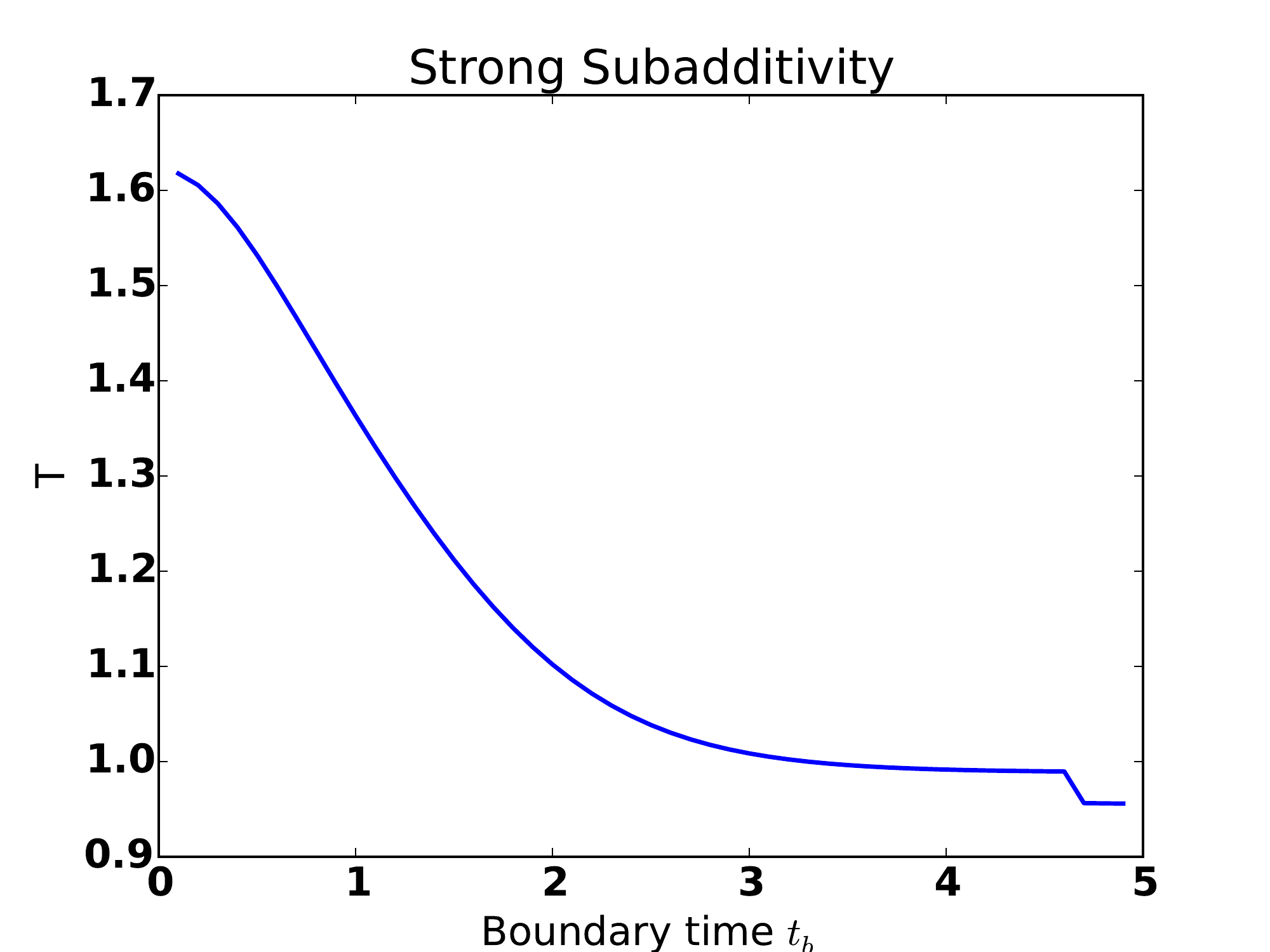}

 \caption{The quantity $T$ plotted vs $t_b$. }\label{fig:SSA}
 \end{subfigure}
 \begin{subfigure}{0.45\textwidth}
 \includegraphics[width=\textwidth]{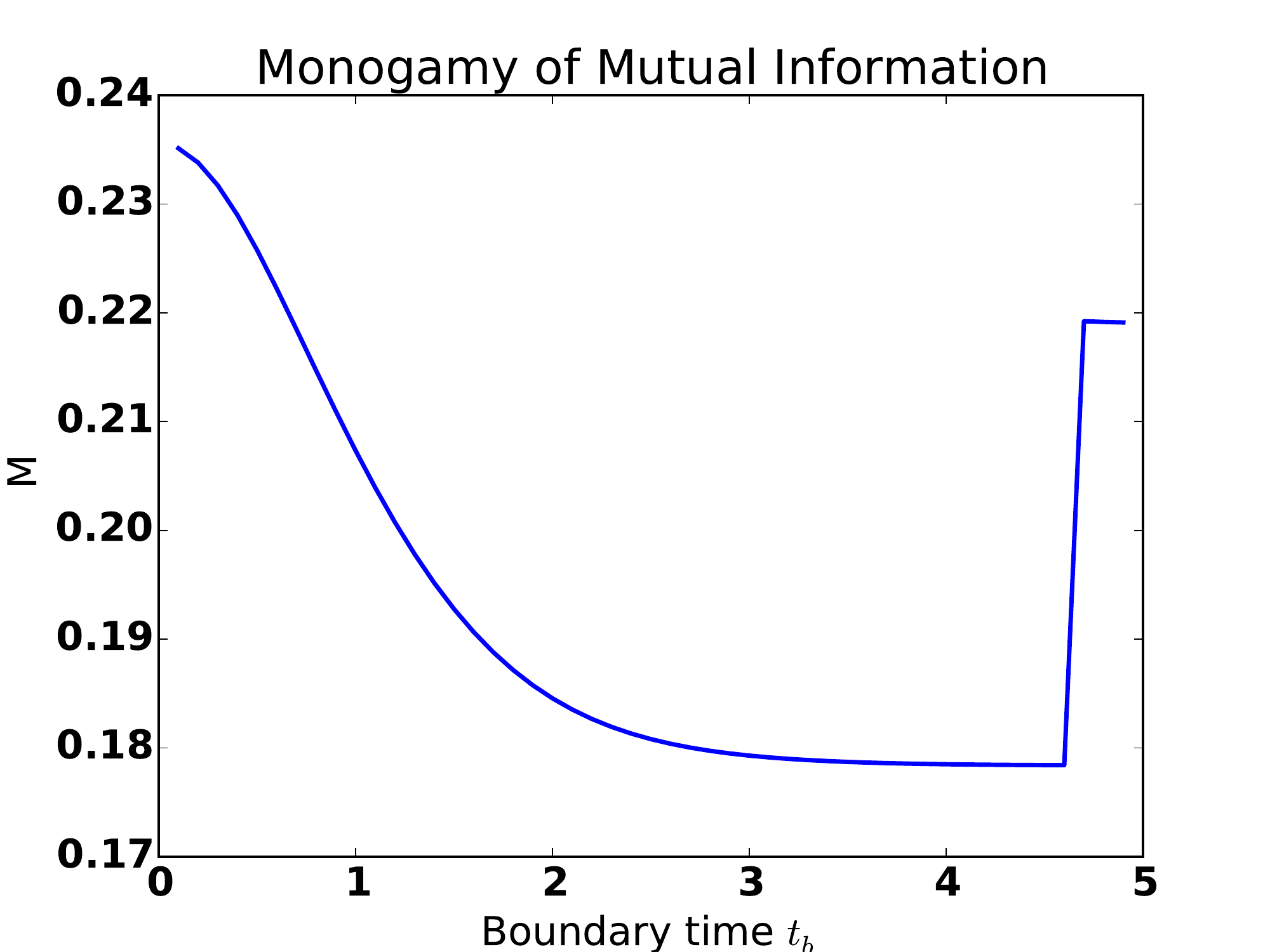}

 \caption{The quantity $M$ plotted vs $t_b$. }\label{fig:MMI}
 \end{subfigure}
 \caption{The quantities $T$ and $M$ plotted vs $t_b$. We see that $T>0$ and $M>0$ so strong subadditivity and monogamy of mutual information are always satisfied. The jumps at late times are due to numerical errors.}
\end{figure}

We begin by testing strong subadditivity. We consider adjacent three regions $A,B,C$, all at constant time $t_b$. We choose $\ell_A=2, \ell_B=4, \ell_C=2$. We then consider the quantity
$$T(A,B,C) \equiv 4G_N[ S(AB)+S(BC)-S(ABC)-S(B)]$$
as a function of the boundary time $t_b$. Strong subadditivity will be satisfied if and only if $T \geq 0$. $S(AB)$ is given by the length of a geodesic with $\ell_x=6$ over $4G_N$. $S(BC)$ is identical. $ABC$ is an interval of length 8, while $B$ is an interval of length 4. Using this information, we can plot $T$ as a function of $t_b$. This is done in Figure~\ref{fig:SSA}. We see that $T>0$, so strong subadditivity is always satisfied.

We do the same calculation for monogamy of mutual information. Define 
\begin{multline}
M \equiv 4G_N[I(A:BC)-I(A:B)-I(A:C)] \\ = 4G_N [S(A)+S(BC)-S(ABC) - S(A) -S(B)  +S(AB) -S(A)-S(C)+S(AC) ] \\ = 4G_N [-S(A)+S(BC)-S(ABC) -S(B)  +S(AB) -S(C)+S(AC) ].
\end{multline}
We plot this quantity as a function of $t_b$ in Figure~\ref{fig:MMI}. Again, since $M>0$, we see that monogamy of mutual information is always satisfied. 

\begin{figure}

\centering

 \includegraphics[width=0.7\textwidth]{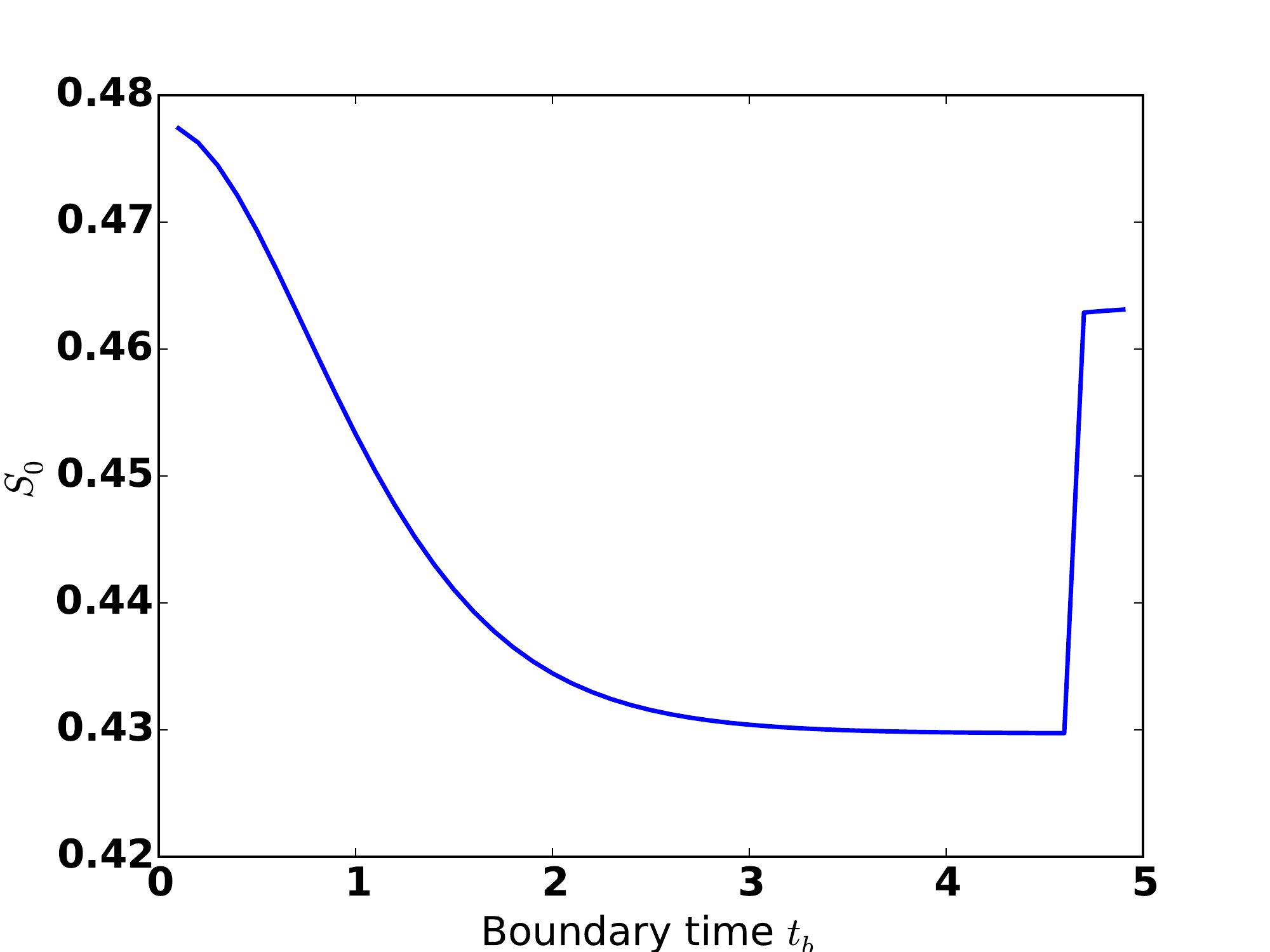}


 \caption{The quantity $S_0$ plotted vs $t_b$. We see that $S_0>0$ so the corresponding inequality is always satisfied. The jump at late times is due to numerical errors.}\label{fig:S0}
\end{figure}

Next, we consider five regions. There are several inequalities that are valid for the RT formula for five regions. These bound the holographic entropy cone. For example, we have that 
$$S(A|BC) + S(B|CD) + S(C|DE) + S(D|EA) + S(E|AB) \geq S(ABCDE) $$
(see \cite{Cone}). Here $S(X|Y) \equiv S(XY)-S(Y)$ is the conditional entropy. 
To test this in the non-static case, consider 5 regions $A,B,C,D,E$, all of which are constant time intervals on the boundary. Take $\ell_A=\ell_C=\ell_E=2$, $\ell_B=\ell_D=4$. Then define the quantity 
$$S_0 \equiv 4G_N[S(A|BC) + S(B|CD) + S(C|DE) + S(D|EA) + S(E|AB) - S(ABCDE)]. $$
We plot this as a function of $t_b$. See Figure~\ref{fig:S0}. We see that $S_0>0$ so this inequality is always satisfied in this non-static case. 

There are several more inequalities for the holographic five-region case \cite{Cone}. For example,
\begin{multline}
2S(ABC)+S(ABD)+S(ABE)+S(ACD)+S(ADE)+S(BCE)+S(BDE) \geq  \\ S(AB)+S(ABCD) + S(ABCE) + S(ABDE) + S(AC) + S(AD) + S(BC) + S(BE) + S(DE),
\end{multline}

\begin{multline}
S(ABE)+S(ABC)+S(ABD)+S(ACD)+S(ACE)+S(ADE)+S(BCE)+S(BDE)+S(CDE) \geq \\ S(AB) + S(ABCE) +S(ABDE) + S(AC) + S(ACDE) + S(AD) + S(BCD) + S(BE) + S(CE) + S(DE),
\end{multline}

\begin{multline}
S(ABC) + S(ABD) + S(ABE) + S(ACD) + S(ACE) + S(BC) + S(DE) \geq  \\ S(AB) +S(ABCD) + S(ABCE) + S(AC) + S(ADE) + S(B) + S(C) + S(D) + S(E),
\end{multline}

\begin{multline}
3S(ABC) + 3S(ABD) + 3S(ACE) + S(ABE) + S(ACD) + S(ADE) + S(BCD) +S(BCE) \\ + S(BDE) + S(CDE) \geq  2S(AB) + 2S(ABCD) + 2S(ABCE) + 2S(AC) +2S(BD) + 2S(CE) + \\ S(ABDE) + S(ACDE) + S(AD) + S(AE) + S(BC) + S(DE).
\end{multline}

\begin{figure}

\centering
\begin{subfigure}{0.45\textwidth}
 \includegraphics[width=\textwidth]{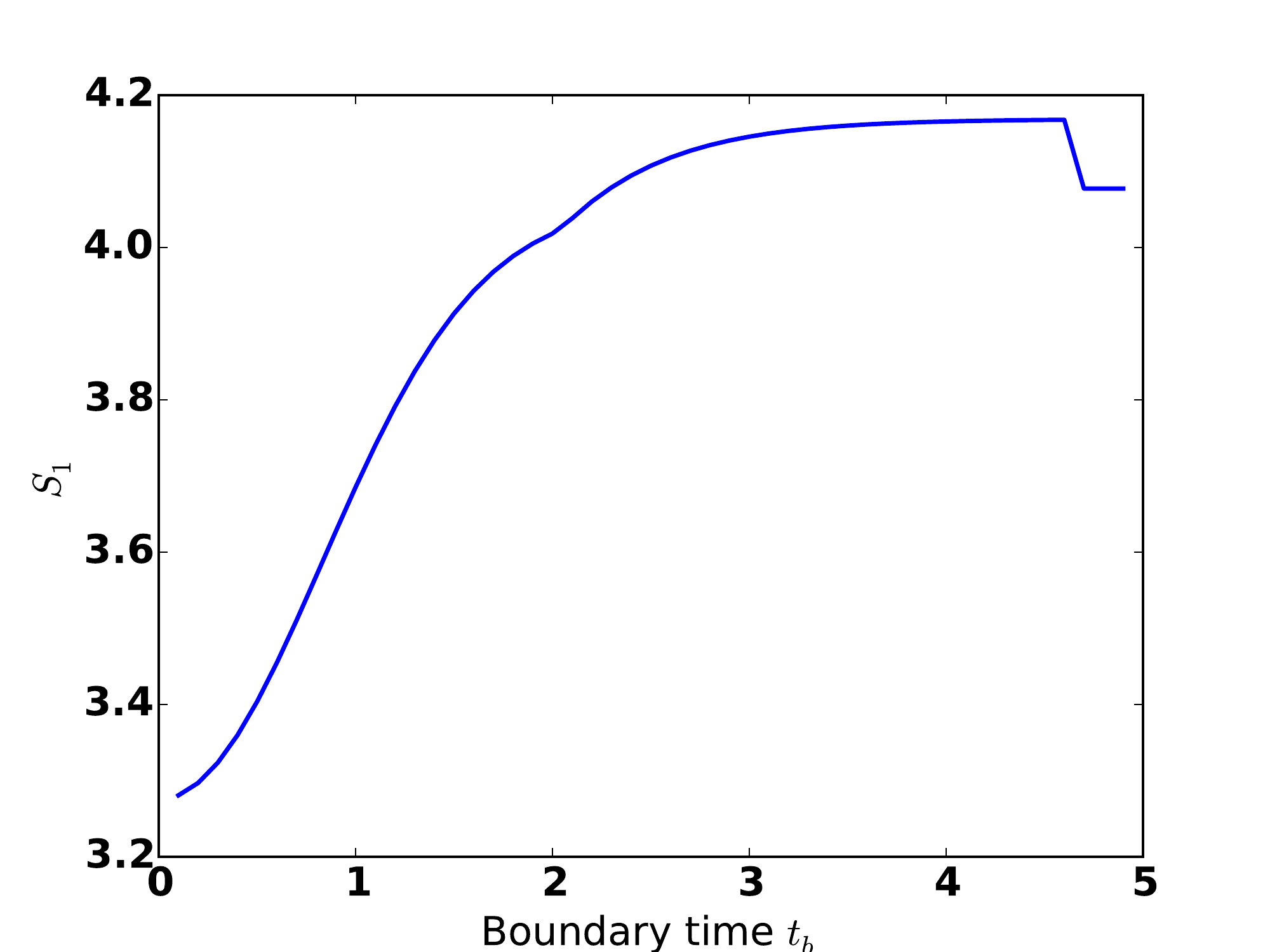}

 \caption{The quantity $S_1$ plotted vs $t_b$. }\label{fig:S1}
 \end{subfigure}
 \begin{subfigure}{0.45\textwidth}
 \includegraphics[width=\textwidth]{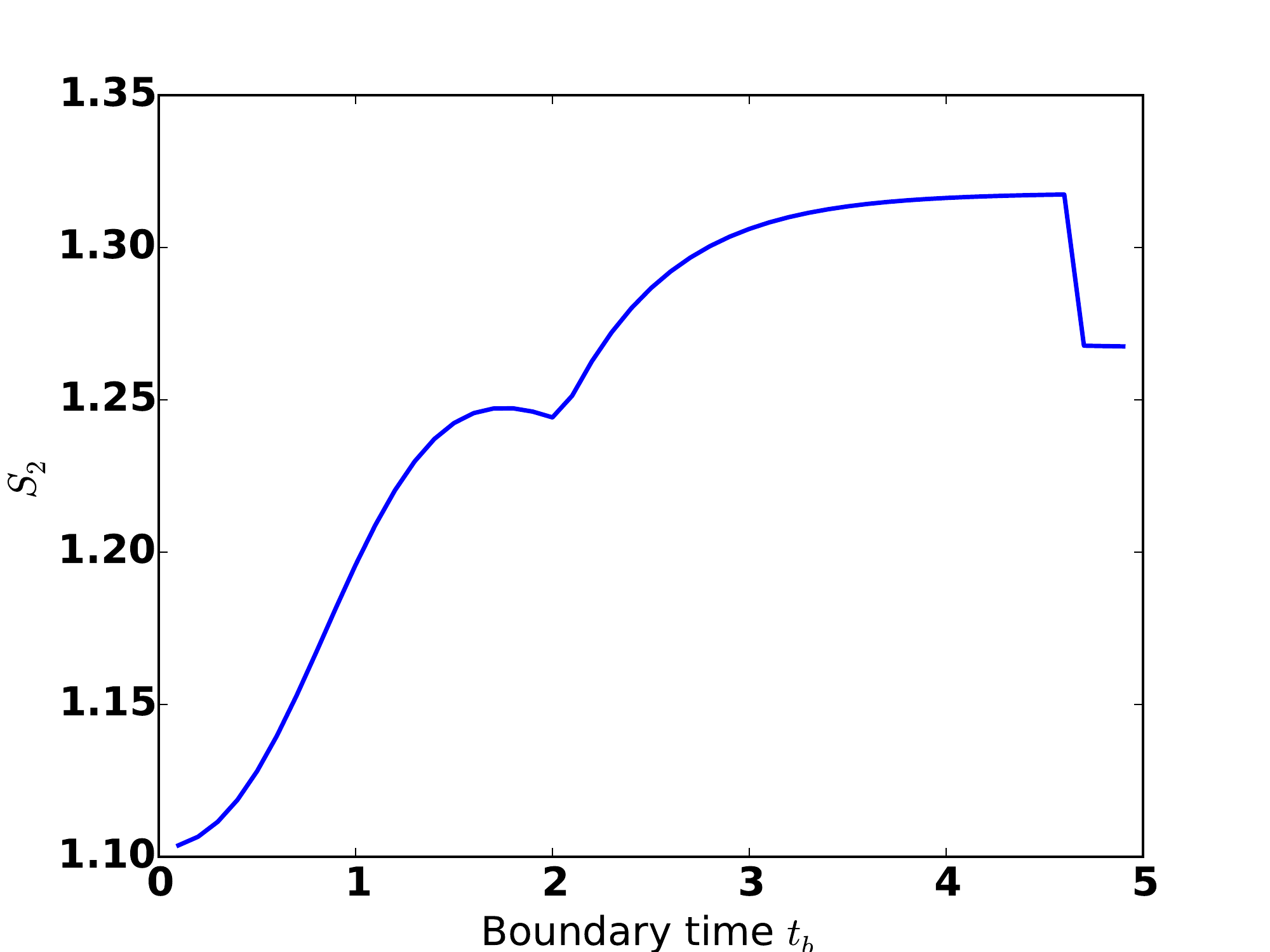}

 \caption{The quantity $S_2$ plotted vs $t_b$. }\label{fig:S2}
 \end{subfigure}
 \begin{subfigure}{0.45\textwidth}
 \includegraphics[width=\textwidth]{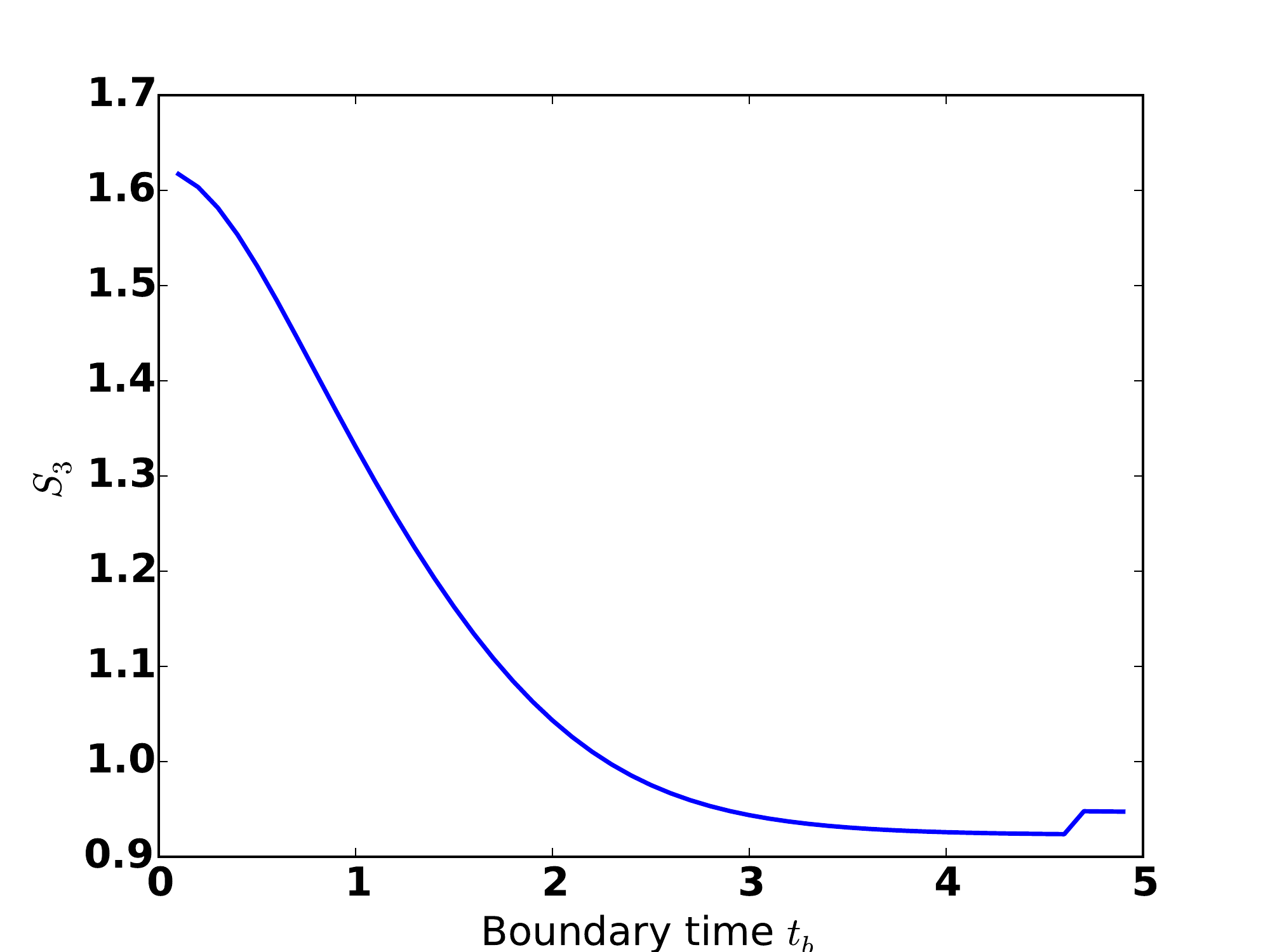}

 \caption{The quantity $S_3$ plotted vs $t_b$. }\label{fig:S3}
 \end{subfigure}
 \begin{subfigure}{0.45\textwidth}
 \includegraphics[width=\textwidth]{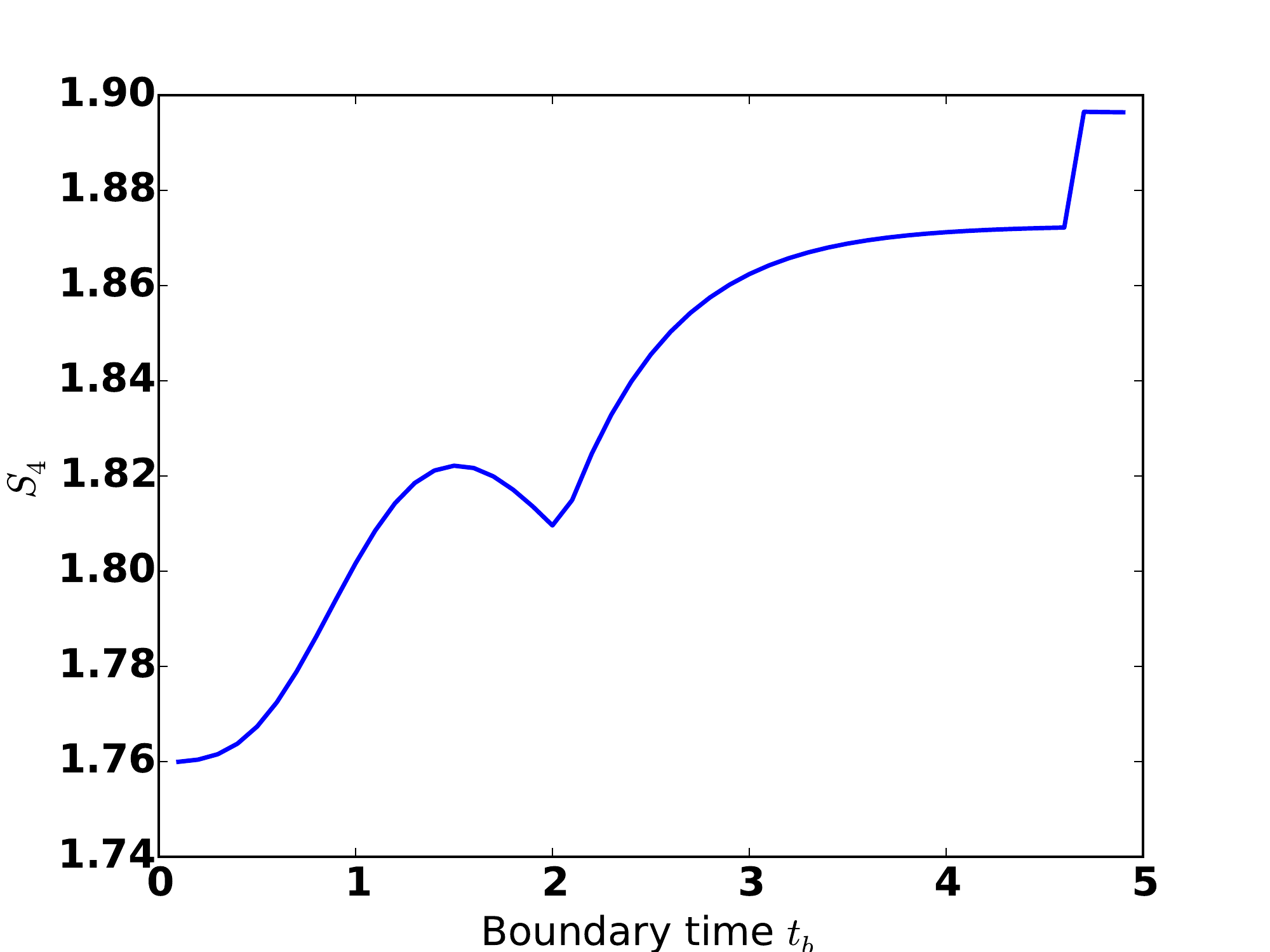}

 \caption{The quantity $S_4$ plotted vs $t_b$. }\label{fig:S4}
 \end{subfigure}
 
 \caption{The quantities $S_i$ ($i=1,2,3,4$) plotted vs $t_b$. We see that $S_i>0$ for all $i$ so the corresponding inequalities given in the main text are always satisfied. The jumps at late times are due to numerical errors.}\label{fig:five}
\end{figure}

For each of these inequalities, we define the quantities $S_i$ to be $4G_N$ times the left-hand side minus $4G_N$ times the right-hand side for $i=1, 2,3,4$. Inequality $i$ will be satisfied if and only if $S_i$ is positive. We plot each of these quantities as functions of $t_b.$ See Figure~\ref{fig:five}. We see that for each $i$, $S_i$ is positive so that the five-region inequalities are all satisfied in this case, even though the spacetime is not static. 

We now consider the case where the interval is not constant-time.

\subsection{Spacelike Intervals with Nonzero $\Delta t$}
\begin{figure}
\centering
 \includegraphics[width=0.35\textwidth]{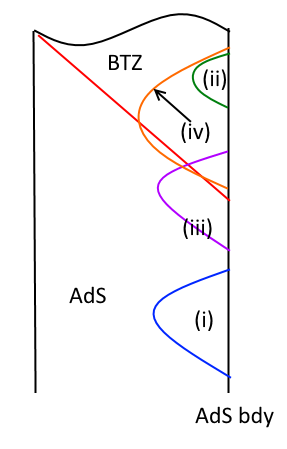}

 \caption{A Penrose diagram of the Vaidya spacetime (the red line represents the shell of null matter), showing the four cases for the general spacelike geodesics. (i) Entirely in the AdS bulk, (ii) entirely in the BTZ bulk, (iii) starts in AdS, crosses into BTZ, and (iv) starts in BTZ, crosses into AdS, and crosses back into BTZ.}\label{fig:Pen_var}
\end{figure}

In this situation, there are four cases: (i) entirely in AdS, (ii) entirely in BTZ, (iii) starts in AdS, crosses into BTZ, (iv) starts in BTZ, crosses into AdS, crosses back into BTZ. See Figure~\ref{fig:Pen_var}. Without loss of generality, suppose $\Delta t \geq 0$. Then the interval is characterized by three parameters, $\Delta x, \Delta t,$ and the starting boundary time of the interval $t_b.$ Again, because the shell is located at $v=0$, if $t_b<0$, the curve is either (i) or (iii). If $t_b \geq 0$, the curve is either (ii) or (iv).

\subsubsection{Geodesics entirely in AdS or BTZ}

We begin with the AdS case. The solution is:
$$r(\tau) = \frac{1}{2}((p_x^2 -E^2)e^{-\tau} + e^\tau),$$
$$t(\tau) = Const - \frac{2E}{p_x^2-E^2+e^{2 \tau}},$$
$$x(\tau) = Const - \frac{2p_x}{p_x^2-E^2+e^{2 \tau}}.$$
This is very similar to the case considered above with $E=0$. We calculate
$$\Delta x = \frac{2 p_x}{p_x^2-E^2}, \Delta t =  \frac{2 E}{p_x^2-E^2}, L_{reg} = -\log ( p_x^2-E^2).$$
We calculated the solution to the BTZ case with $E \neq 0$ above. The solution is 
$$r(\tau)^2 = \frac{1}{4} (e^\tau + B_+ e^{-\tau})(e^\tau + B_- e^{-\tau}),$$
$$t(\tau) =Const+ \frac{1} {2} \log \left ( \frac{A_-+e^{2 \tau}}{A_+ + e^{2 \tau}} \right),$$
$$x(\tau) = Const - \frac{1} {2} \log \left ( \frac{B_-+e^{2 \tau}}{B_+ + e^{2 \tau}} \right),$$
where we have defined the quantities
$$B_{\pm} = (p_x \pm 1)^2 -E^2, A_\pm = p_x^2 -(1 \pm E)^2.$$
This tells us that 
$$ \Delta x = \frac{1}{2} \log \left ( \frac{A_-}{A_+} \right ), \Delta t = -\frac{1}{2} \log \left ( \frac{B_-}{B_+} \right ), L_{reg} = -\frac{1}{2} \log(B_+ B_-).$$
We now turn our attention to the geodesics that cross the shell. 

\subsubsection{Geodesics that start in AdS, end in BTZ}
In this case, the geodesic intersects the shell once, say at coordinate $r_c$. The affine parameter (in the AdS portion) at which the crossing occurs is 
$$ \tau_c = \log (r_c +\sqrt{r_c^2+E_A^2-p_x^2} )$$
Because this occurs at the shell, we need $t_c=t(\tau_c) = 1/r_c,$ which fixes the constant in the equation for $t(\tau)$. The length of the geodesic in AdS is given by 
$$L_A = \log 2R + \log ( r_c +\sqrt{r_c^2 + E_A^2 -p_x^2} )- \log (p_x^2 -E_A^2).$$
Meanwhile, 
$$\Delta x_A = \frac{2 p_x}{p_x^2-E_A^2} - \frac{p_x}{r_c (r_c +\sqrt{r_c^2+E_A^2-p_x^2})}, \Delta t_A = \frac{E_A} {p_x} \Delta x_A.$$
As we calculated above, the value of $E$ in the BTZ portion is given by 
$$E_B = (1-\frac{1}{2 r_c^2} ) E_A-\frac{p_x r_A'}{2 r_c^4}, \text{     } r_A'= \sqrt{\frac{r_c^6}{p_x^2}+\frac{r_c^4 E_A^2}{p_x^2}-r_c^4}.$$
The value of the affine parameter in the BTZ portion of the crossing is 
$$ \alpha_B \equiv \exp(2 \tau_B) = \frac{1}{2} \left [ -(B_+ +B_-) + 4r_c^2+\sqrt{-4B_+ B_-+ (B_+ +B_--4r_c^2)^2} \right ], $$
where $B_\pm$ is as defined above, using the energy $E_B$. Furthermore, we have

$$\Delta x_B= x(\tau = \infty) - x(\tau_B)= -\frac{1} {2} \log \left ( \frac{B_-+\alpha_B}{B_+ + \alpha_B} \right),$$
$$\Delta t_B = \frac{1} {2} \log \left ( \frac{A_-+\alpha_B}{A_+ + \alpha_B} \right),$$
$$L_B = \log 2R - \frac{1}{2} \log \alpha_B.$$
In the AdS region, the time of the boundary crossing is given by $t_c= 1/r_c$. Therefore, the starting time of the interval is 
$$t_b = \frac{1}{r_c} -\Delta t_A.$$
The total (regularized) length of the curve is 
$$ L_{reg} = -\frac{1}{2} \log \alpha_c =\log 2R + \log ( r_c +\sqrt{r_c^2 + E_A^2 -p_x^2} )- \log (p_x^2 -E_A^2),$$
while 
$$\Delta x = \Delta x_A + \Delta x_B, \Delta t = \Delta t_A  + \Delta t_B.$$

\subsubsection{Geodesics that start in BTZ, cross into AdS, end in BTZ}

Finally we consider the geodesics that start in the BTZ bulk (so that $t_b=0$), cross over into the AdS bulk, and then cross back to the BTZ bulk. These geodesics cross the shell twice, say at $r_1$ and $r_2$, with $r_1>r_2$. If the part of the geodesic in AdS has $E_A$, then the length of the AdS portion is given by 
$$L_A = \tau_1 - \tau_2 = \log ( r_1 +\sqrt{r_1^2 + E_A^2 -p_x^2} )-\log ( r_2 +\sqrt{r_2^2 + E_A^2 -p_x^2} ).$$
Also,
$$ \Delta t_A = \frac{1}{r_2} - \frac{1}{r_1}, \Delta x_A = \frac{p_x}{E_A} \Delta t_A.$$
We now consider the BTZ portions of the geodesics. Consider the upper BTZ arc of the geodesic. The shell is at $v=0$, so since $r_1 > r_2$, $r_1'<0$. Thus, we obtain
$$r_A^{1 \prime}= -\sqrt{\frac{r_1^6}{p_x^2}+\frac{r_1^4 E_A^2}{p_x^2}-r_1^4},$$
and 
$$E_{B1} = (1-\frac{1}{2 r_1^2} ) E_A-\frac{p_x r_A^{1 \prime}}{2 r_1^4}.$$
We know $r$ $x$ and $t$ as functions of $\tau$ for the BTZ curve for these values of the conserved momenta. We can numerically solve for $\tau_B^1$ when $r_{B1}$ is equal to $r_1$. Because $t_{B1} (\tau_B^1 )$ has to be equal to $\tanh^{-1} (1/r_1)$, this fixes the constant. We then compute 
$$ \Delta x_{B1} = x_{B1} (\tau_\infty) - x_{B1} (\tau_B^1),    \Delta x_{B1} = x_{B1} (\infty) - x_{B1} (\tau_B^1), L_{B1} = \tau_\infty - \tau_B^1.$$
We repeat this procedure for the bottom BTZ arc. For $r_2$, however, $r_A^{2 \prime}>0$, so that 
$$r_A^{1 \prime}= \sqrt{\frac{r_2^6}{p_x^2}+\frac{r_2^4 E_A^2}{p_x^2}-r_2^4},$$
which means
$$E_{B2} = (1-\frac{1}{2 r_2^2} ) E_A-\frac{p_x r_A^{1 \prime}}{2 r_2^4}.$$
We then follow the same procedure to compute $\Delta x_{B2}, \Delta t_{B2}$ and $L_{B2}$. The totals are, of course, 
$$\Delta x = \Delta x_A + \Delta x_{B1} + \Delta x_{B2}, \Delta t = \Delta t_A + \Delta t_{B1} + \Delta t_{B2}.$$
To obtain the regularized geodesic length, we have to subtract off the usual UV-divergent term: 
$$L_{reg} = L_A +L_{B1} + L_{B2} -2 \log ( 2 R).$$

The boundary time of the start point of the interval is given by 
$$t_b = \frac{1}{r_1} - \Delta t_{B1}.$$
If we are given $r_1$ and $E_A$, we can calculate $r_2$ as follows. In the AdS region, we can solve for the value of $\tau$ at which $r$ is equal t o$r_1$. We know that $t$ evaluated at this value is $1/r_1$, which fixes the value of the integration constant in the $t(\tau)$ function. We then find the other value of $\tau$ for which the function $v(\tau)= r(\tau) - \frac{1}{t(\tau)}$ vanishes. Evaluating the function $r(\tau)$ at this value then gives us $r_2$. Therefore, the geodesic is specified by three parameters: $E_A$, $p_x$, and $r_1$. From these we can calculate the starting time $t_b$ and the values of $\Delta x$ and $\Delta t$. For values of $t_b$, $\Delta x$ and $\Delta t$, we can numerically find the corresponding values of $E_A$, $p_x,$ and $r_1$, and then use these to calculate the geodesic lengths. 

\subsection{Testing Entropy Inequalities}

\begin{figure}
\centering
 \includegraphics[width=0.7\textwidth]{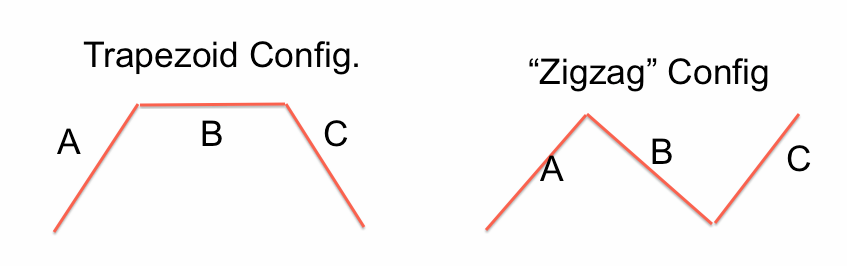}

 \caption{The covariant "trapezoid" and "zigzag" configurations. In both cases, each of the components ($A, B$, and $C$) have $\Delta x = 1.$}\label{fig:Cov_ABC}
\end{figure}

\begin{figure}

\centering
\begin{subfigure}{0.45\textwidth}
 \includegraphics[width=\textwidth]{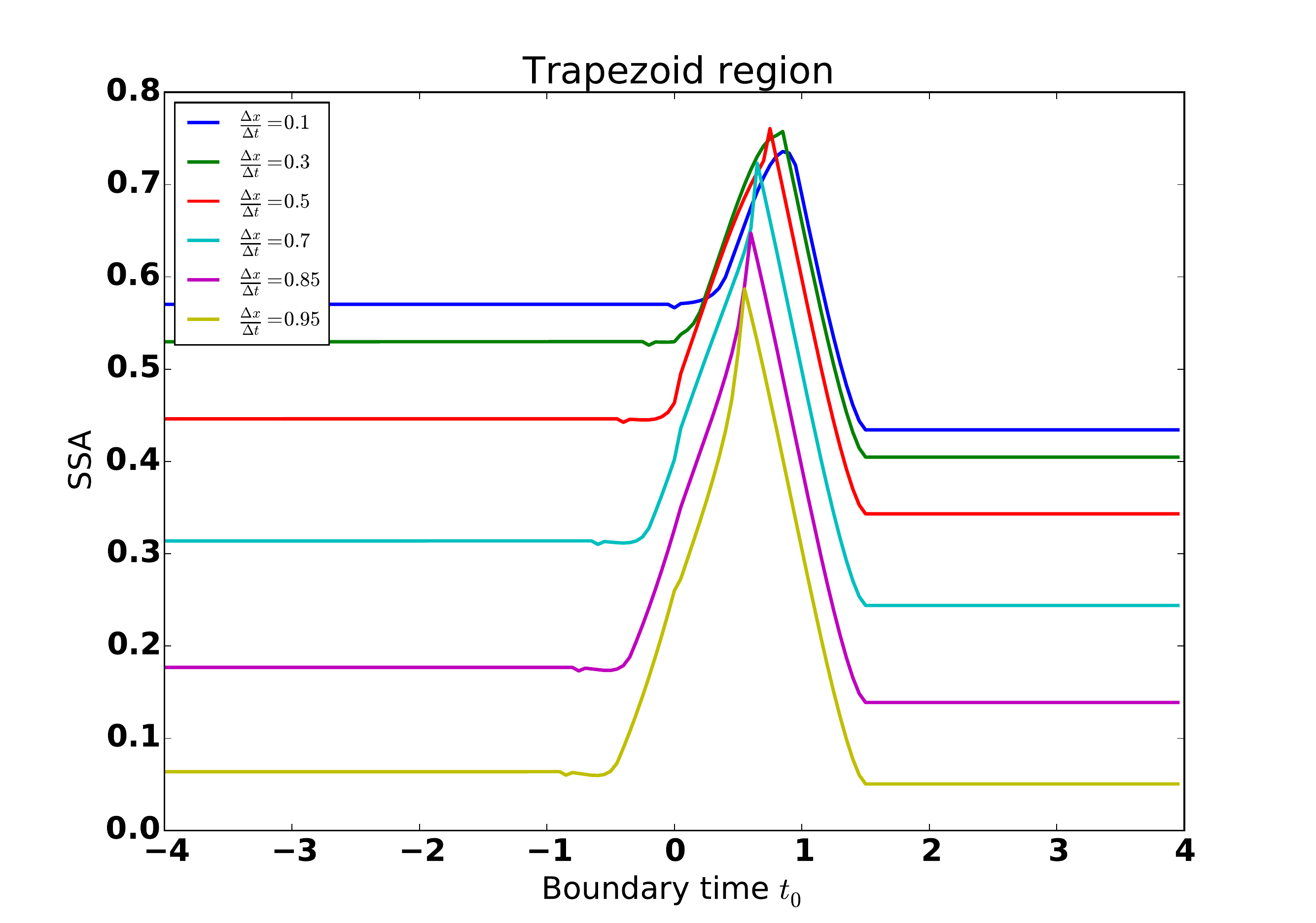}

 \caption{$S(AB) +S(BC) -S(B) -S(ABC)$ versus $t_b$ plotted for the trapezoid configuration. It is always positive so strong subadditivity holds.. }\label{fig:SSA_Tr}
 \end{subfigure}
 \begin{subfigure}{0.45\textwidth}
 \includegraphics[width=\textwidth]{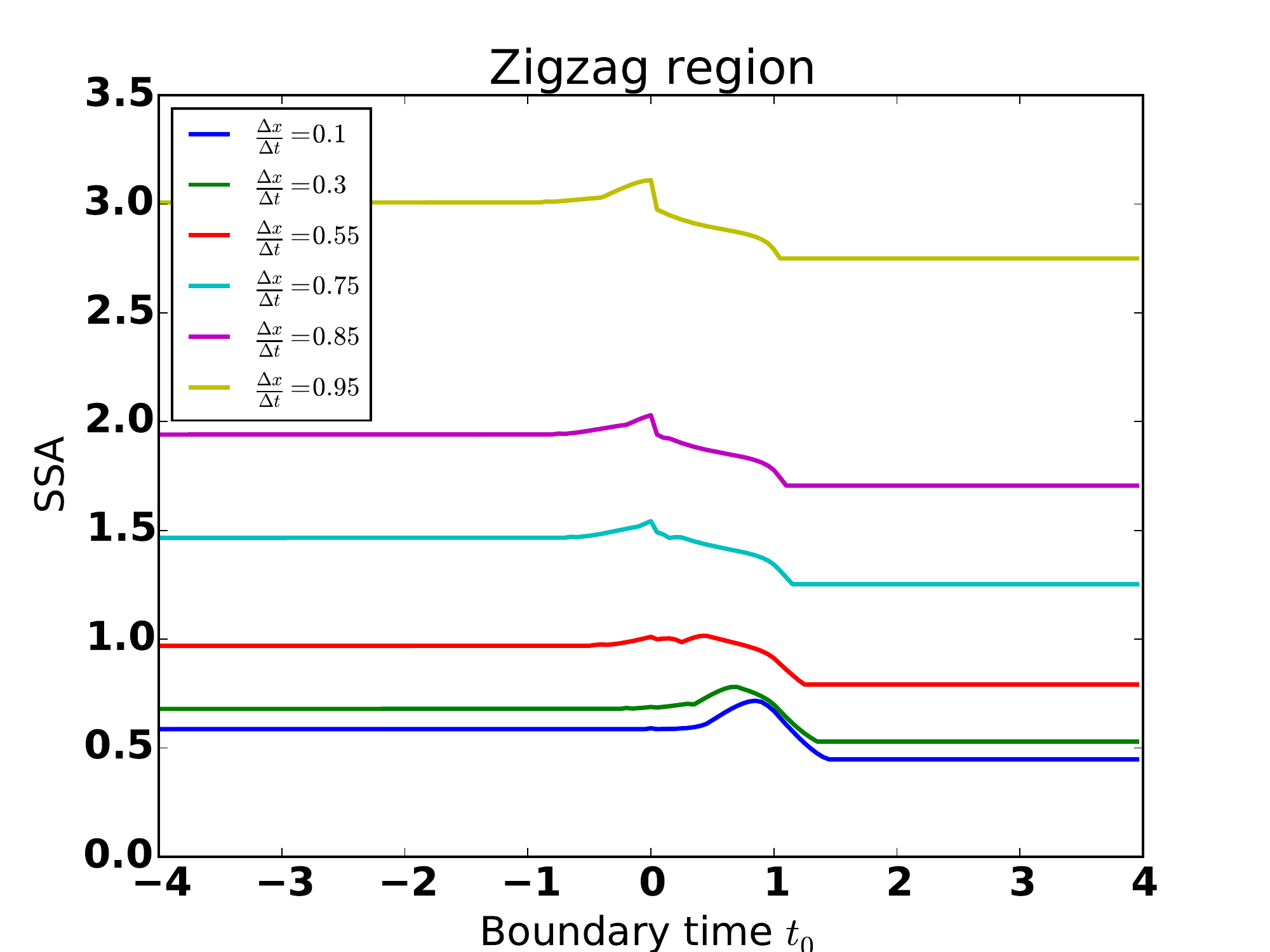}

 \caption{ $S(AB) +S(BC) -S(B) -S(ABC)$ versus $t_b$ plotted for the zigzag configuration. It is always positive so strong subadditivity holds.}\label{fig:SSA_Z}
 \end{subfigure}
 \caption{Strong subadditivity is verified for the trapezoid and zigzag configurations.}
\end{figure}

To find the geodesic length for a given set of parameters, we proceed as follows. If $t_b<0$, we solve for the AdS geodesic. We then calculate $v (\tau) = t(\tau)-\frac{1}{r(\tau)}$. If $v$ never crosses 0, the geodesic is entirely in the AdS bulk. If it crosses $0$, then the geodesic has a portion in the BTZ spacetime. We then numerically find the values of $r_c, p_x, E_A$ that correspond to the given values of $\Delta x, \Delta t, t_b$. We then substitute these results back into our formula for geodesic length. Similarly, if $t_b\geq 0$, we calculate the pure BTZ solution, and see if $v(\tau)=t(\tau) - \tanh^{-1}(\frac{1}{r(\tau)})$ is ever negative, there is a component in the AdS bulk. We numerically find the values of $r_1,E_A, p_x$ that correspond to the values of $\Delta x, \Delta t, t_b$ and use these to find the geodesic length.

Finally, we are ready to test the entropy inequalities for regions that are not purely spacelike. We begin by testing strong sub-additivity. We test two cases, the "trapezoidal" case, and the "zigzag" case; both of these are shown in Figure~\ref{fig:Cov_ABC}. We plot the quantity
$$S(AB) +S(BC) -S(B) -S(ABC)$$
(times $4 G_N$) for these regions as a function of the boundary start time of the region $A$, for a variety of values of $\Delta t/ \Delta x$, fixing $\Delta x =1.$ We show the results in Figures~\ref{fig:SSA_Tr} and \ref{fig:SSA_Z}. These curves show that strong sub-additivity is obeyed for these regions. 

\begin{figure}

\centering
 \includegraphics[width=\textwidth]{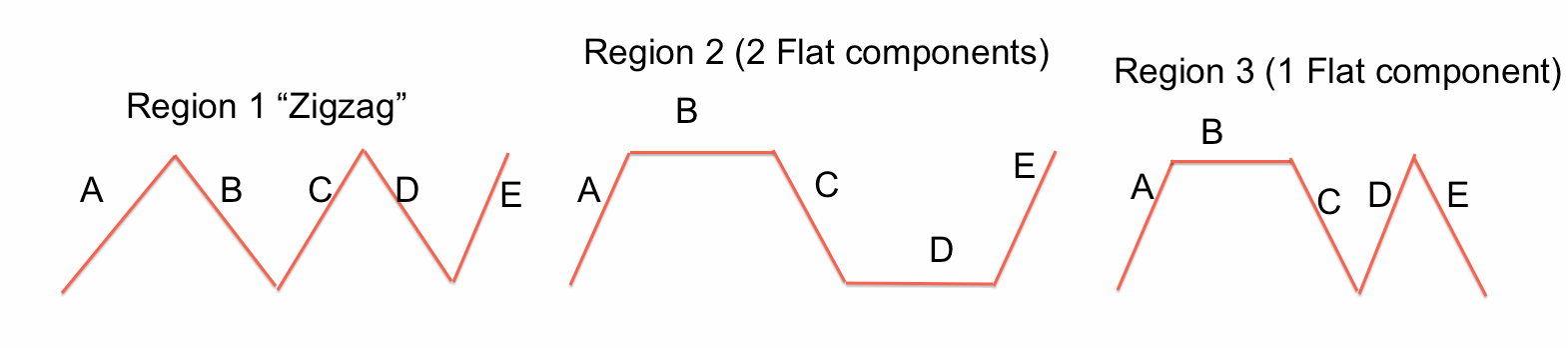}

 \caption{The three configurations we consider. Each of the components has its $\Delta x$ fixed to be 1.}\label{fig:regions}
\end{figure}

\begin{figure}

\centering

\begin{subfigure}{0.45\textwidth}
 \includegraphics[width=\textwidth]{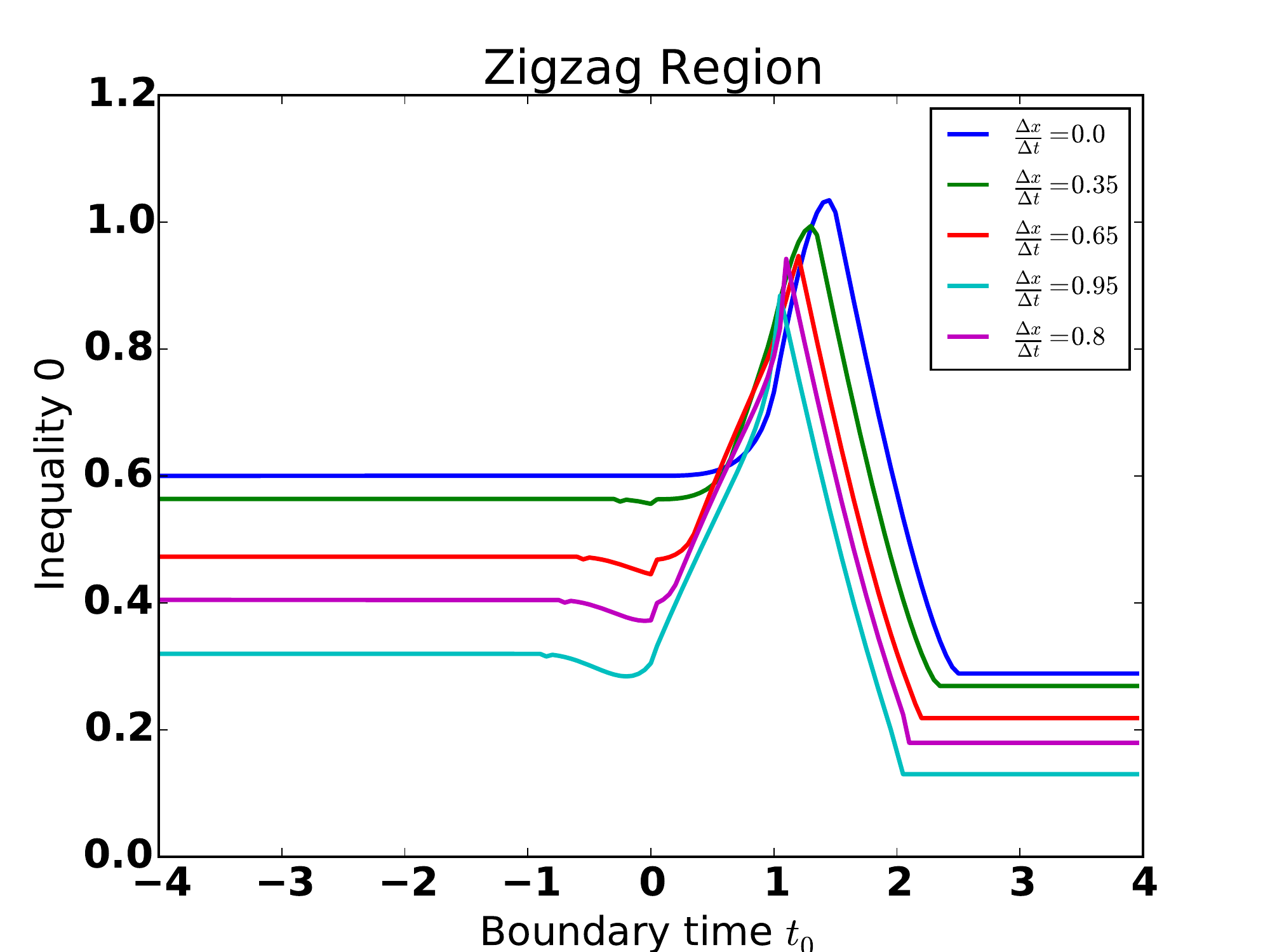}

 \caption{Inequality 0 plotted vs $t_b$. }
 \end{subfigure}

\begin{subfigure}{0.45\textwidth}
 \includegraphics[width=\textwidth]{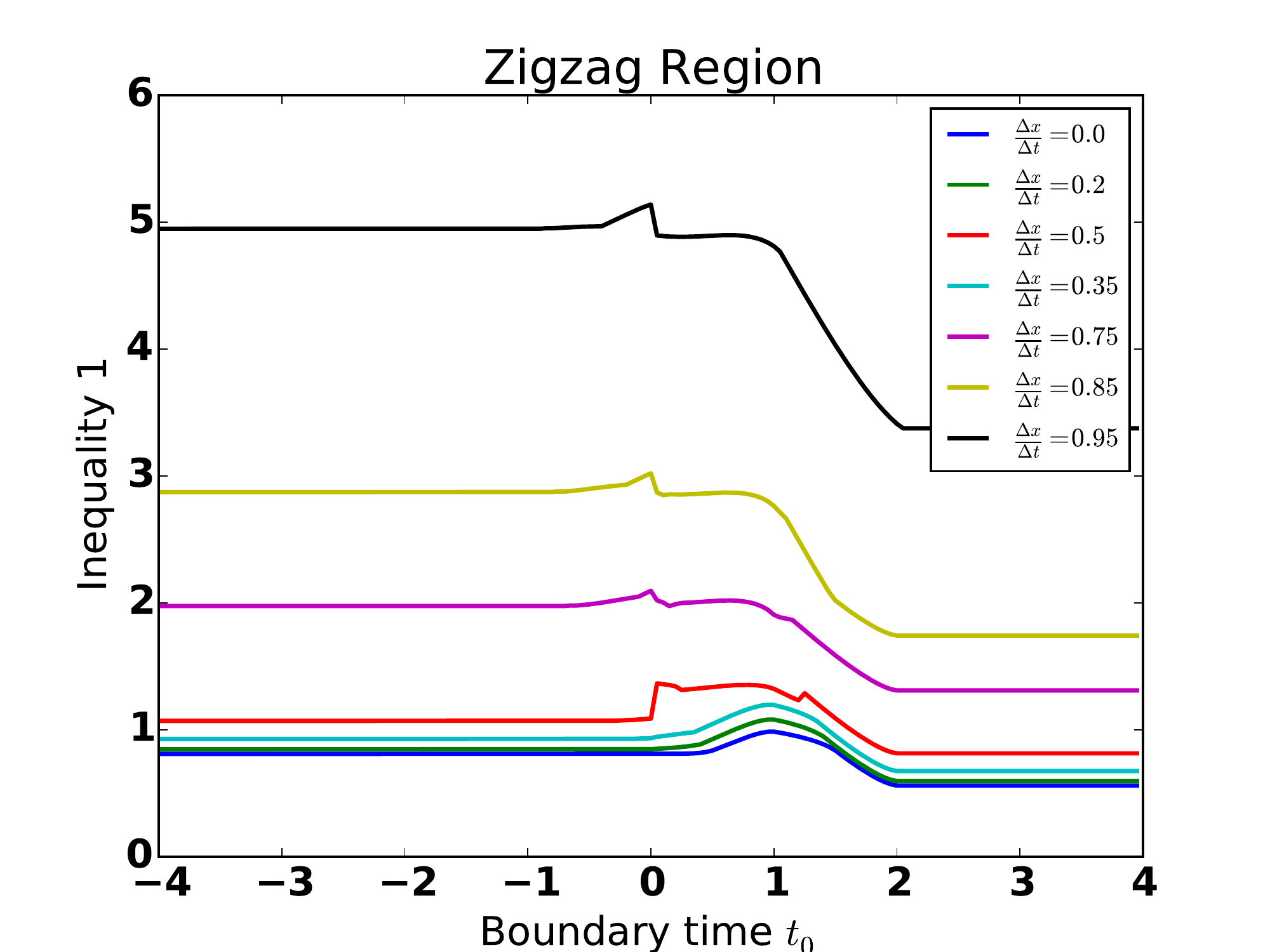}

 \caption{Inequality 1 plotted vs $t_b$. }
 \end{subfigure}
 \begin{subfigure}{0.45\textwidth}
 \includegraphics[width=\textwidth]{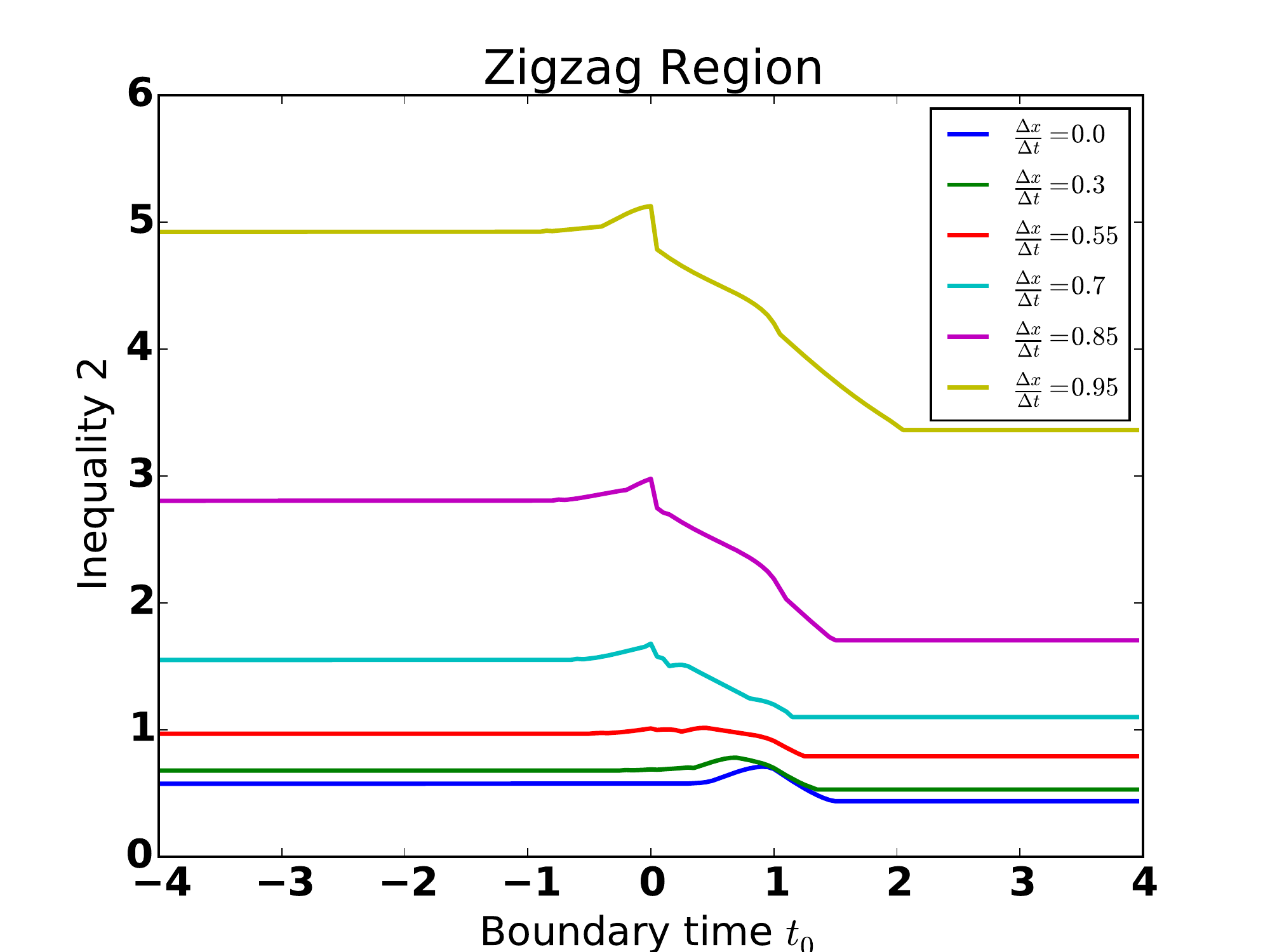}

 \caption{Inequality 2 plotted vs $t_b$. }
 \end{subfigure}
 \begin{subfigure}{0.45\textwidth}
 \includegraphics[width=\textwidth]{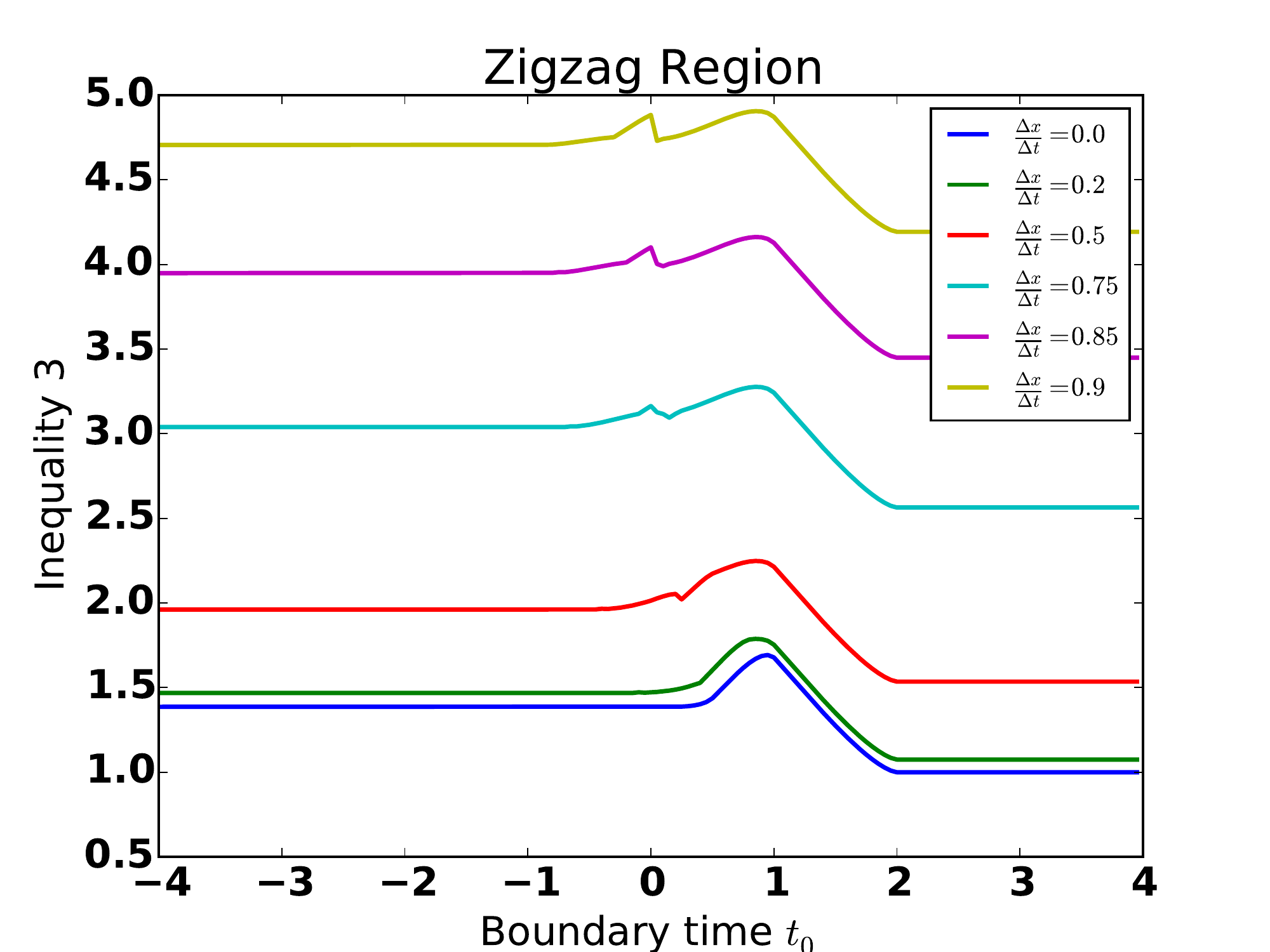}

 \caption{Inequality 3 plotted vs $t_b$. }
 \end{subfigure}
 \begin{subfigure}{0.45\textwidth}
 \includegraphics[width=\textwidth]{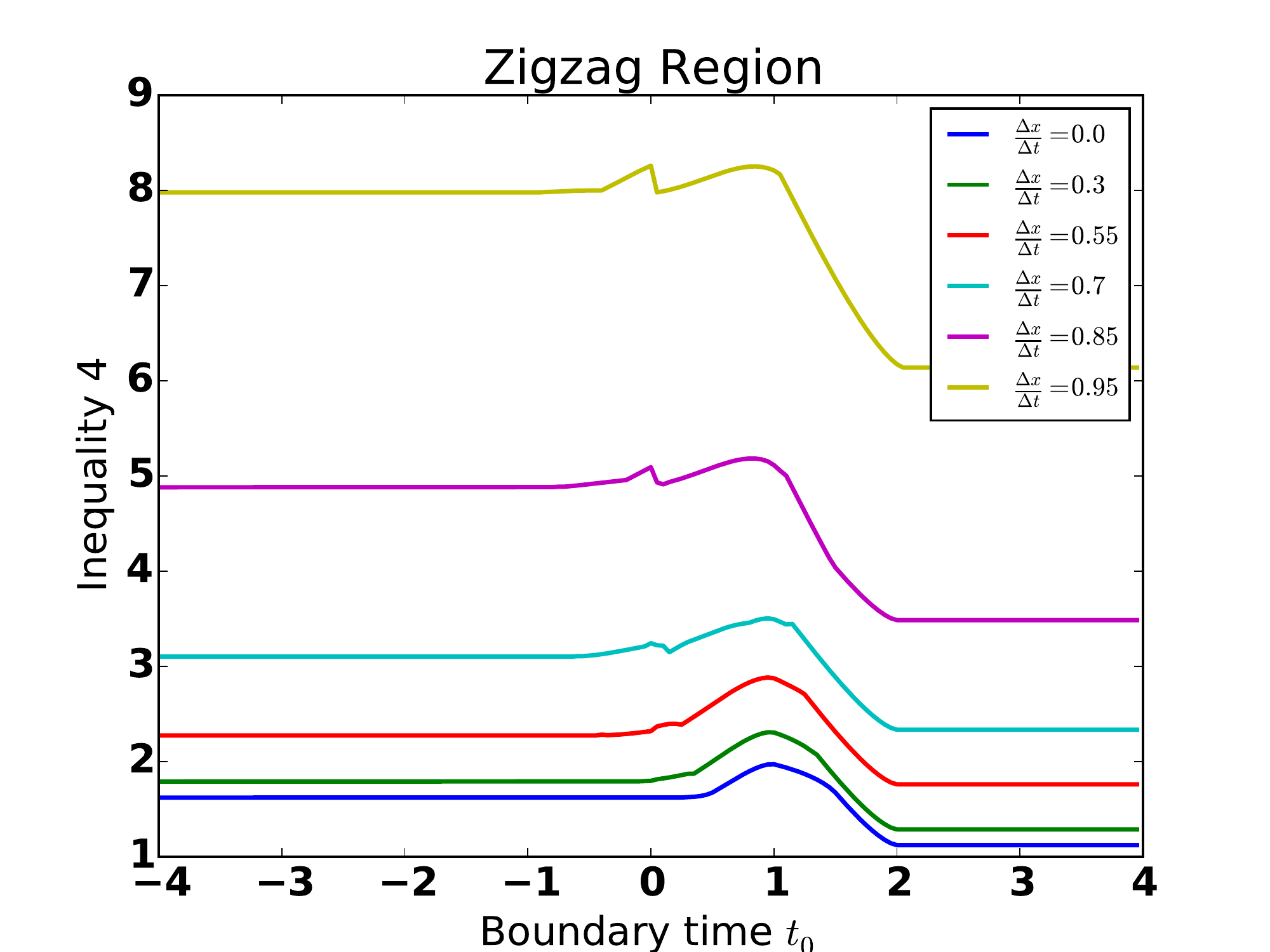}

 \caption{Inequality 4 plotted vs $t_b$. }
 \end{subfigure}
 
 \caption{The inequalities plotted vs $t_b$ for the zigzag region for a variety of values of $\Delta t / \Delta x$. We see that the inequalities are all satisfied.}\label{fig:five_Z}
\end{figure}

\begin{figure}

\centering

\begin{subfigure}{0.45\textwidth}
 \includegraphics[width=\textwidth]{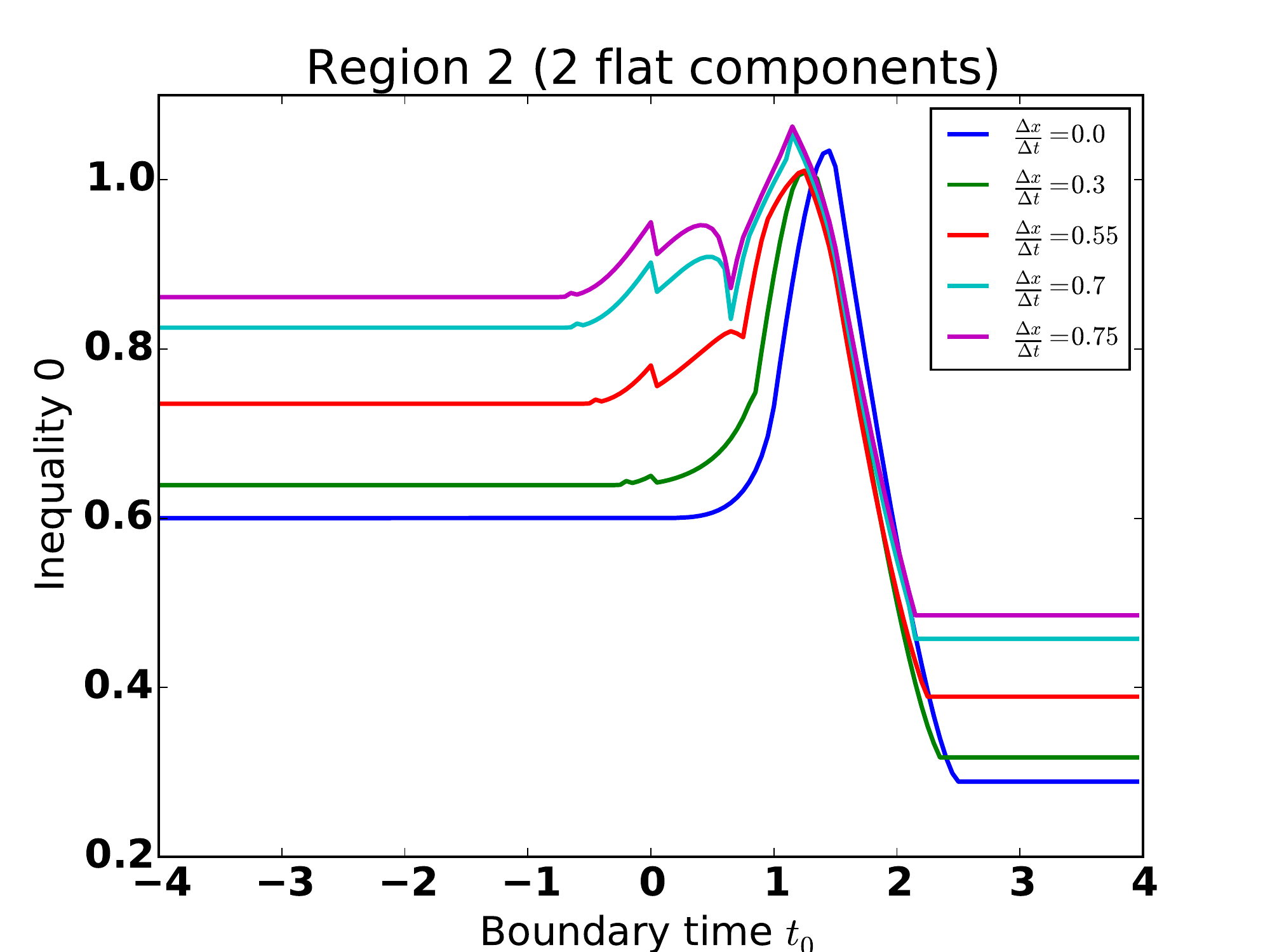}

 \caption{Inequality 0 plotted vs $t_b$. }
 \end{subfigure}

\begin{subfigure}{0.45\textwidth}
 \includegraphics[width=\textwidth]{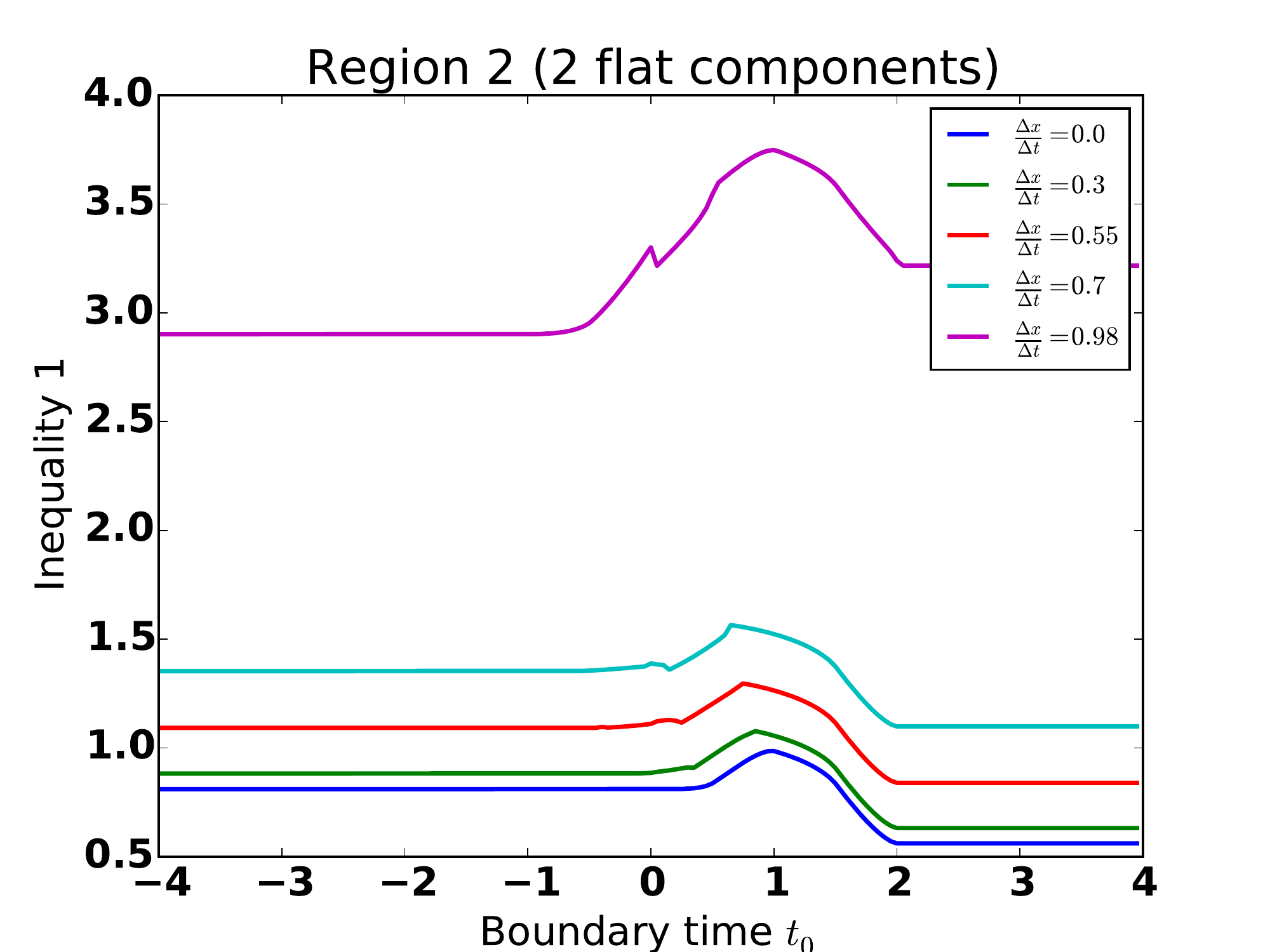}

 \caption{Inequality 1 plotted vs $t_b$. }
 \end{subfigure}
 \begin{subfigure}{0.45\textwidth}
 \includegraphics[width=\textwidth]{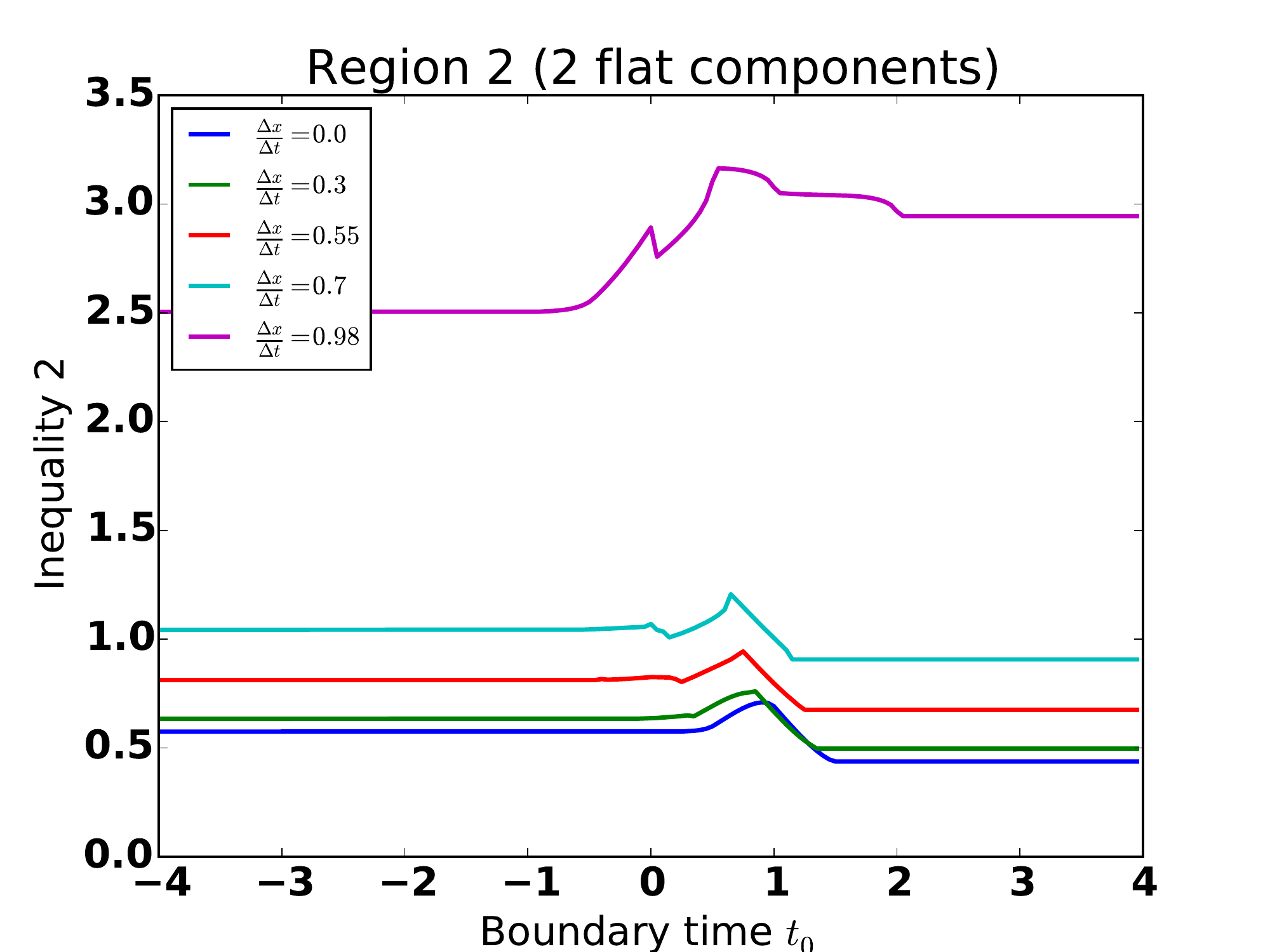}

 \caption{Inequality 2 plotted vs $t_b$. }
 \end{subfigure}
 \begin{subfigure}{0.45\textwidth}
 \includegraphics[width=\textwidth]{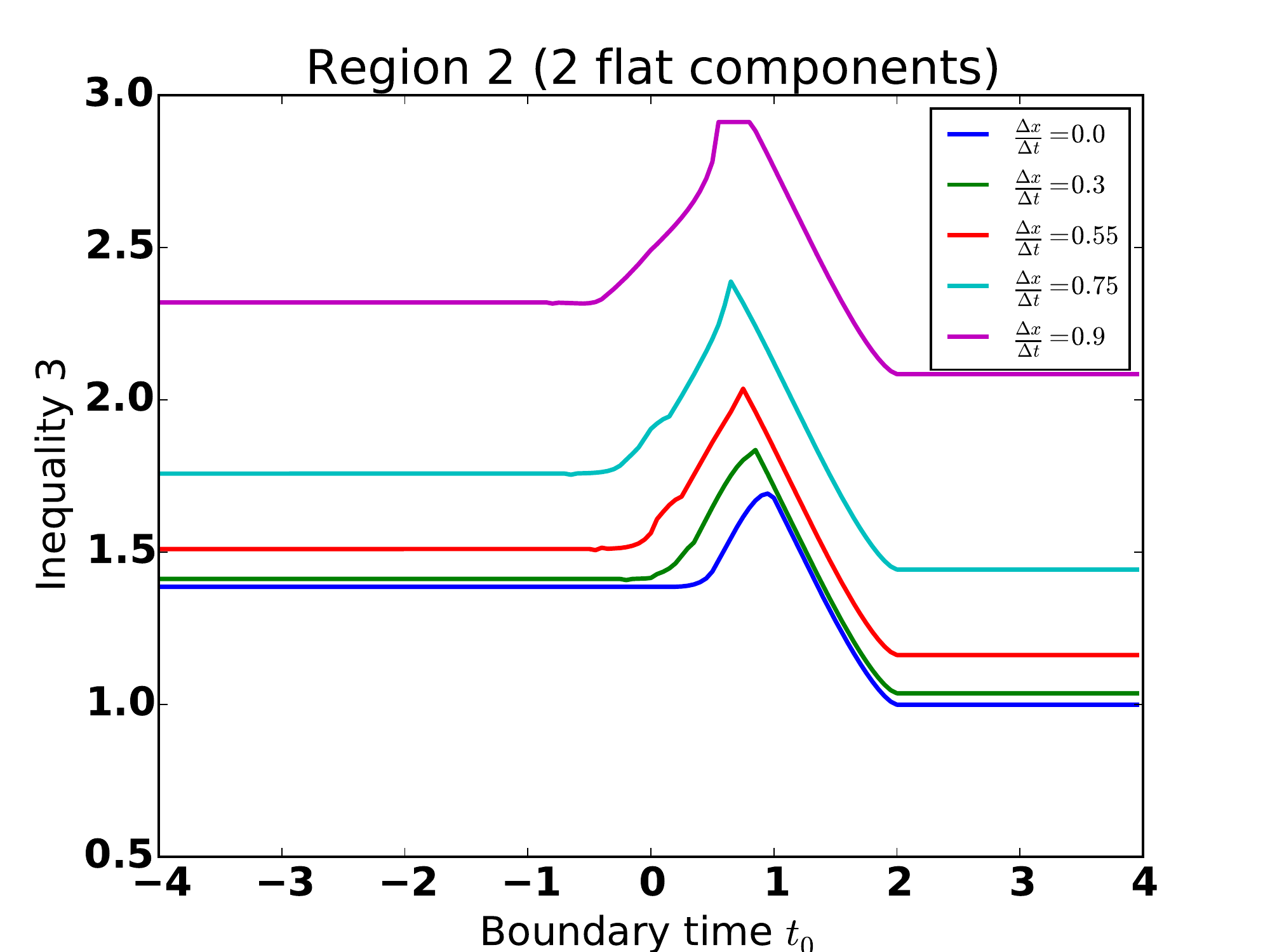}

 \caption{Inequality 3 plotted vs $t_b$. }
 \end{subfigure}
 \begin{subfigure}{0.45\textwidth}
 \includegraphics[width=\textwidth]{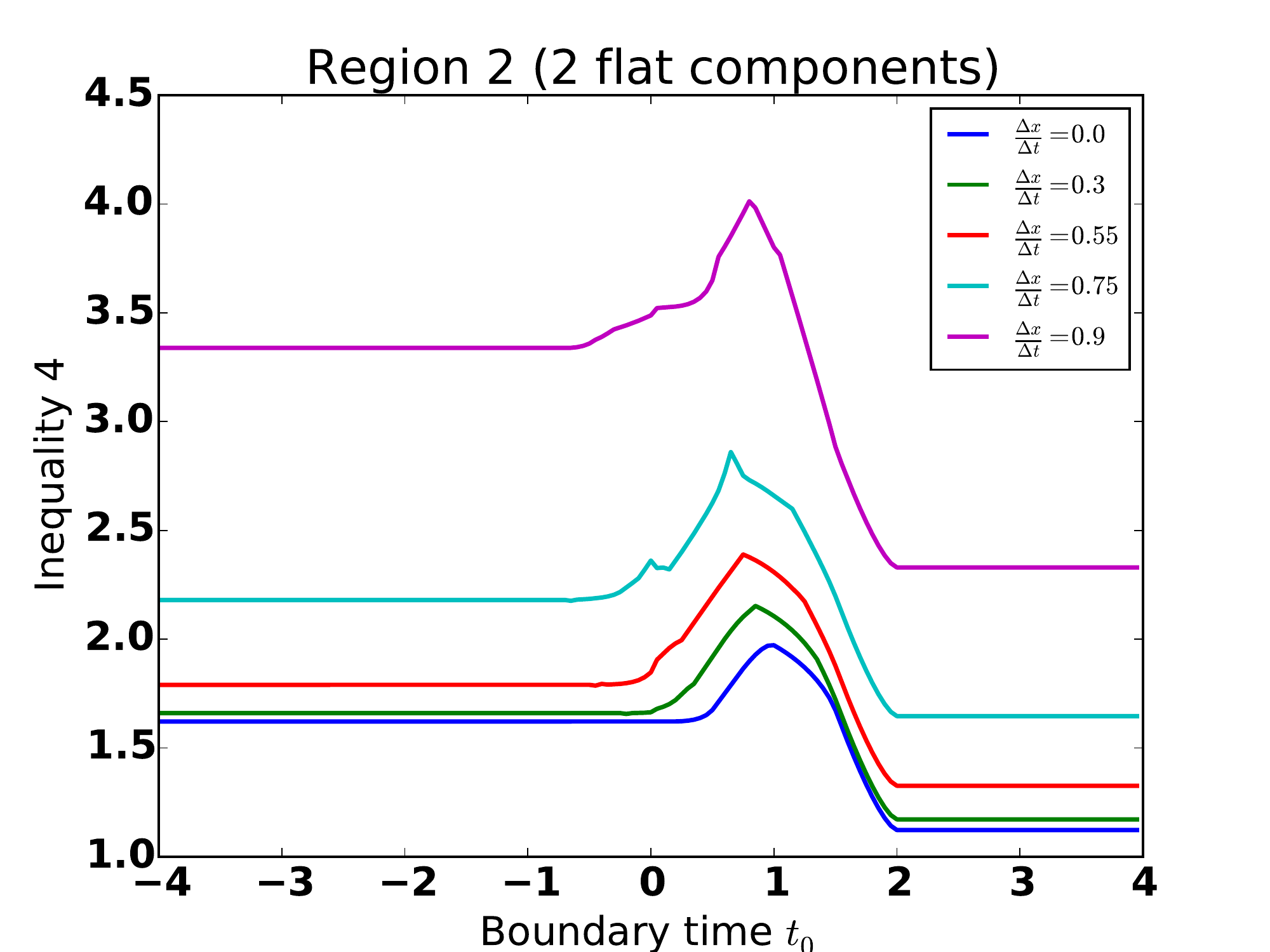}

 \caption{Inequality 4 plotted vs $t_b$. }
 \end{subfigure}
 
 \caption{The inequalities plotted vs $t_b$ for the Region 2 (with 2 flat components) for a variety of values of $\Delta t / \Delta x$. We see that the inequalities are all satisfied.}\label{fig:five_flat2}
\end{figure}

\begin{figure}

\centering

\begin{subfigure}{0.45\textwidth}
 \includegraphics[width=\textwidth]{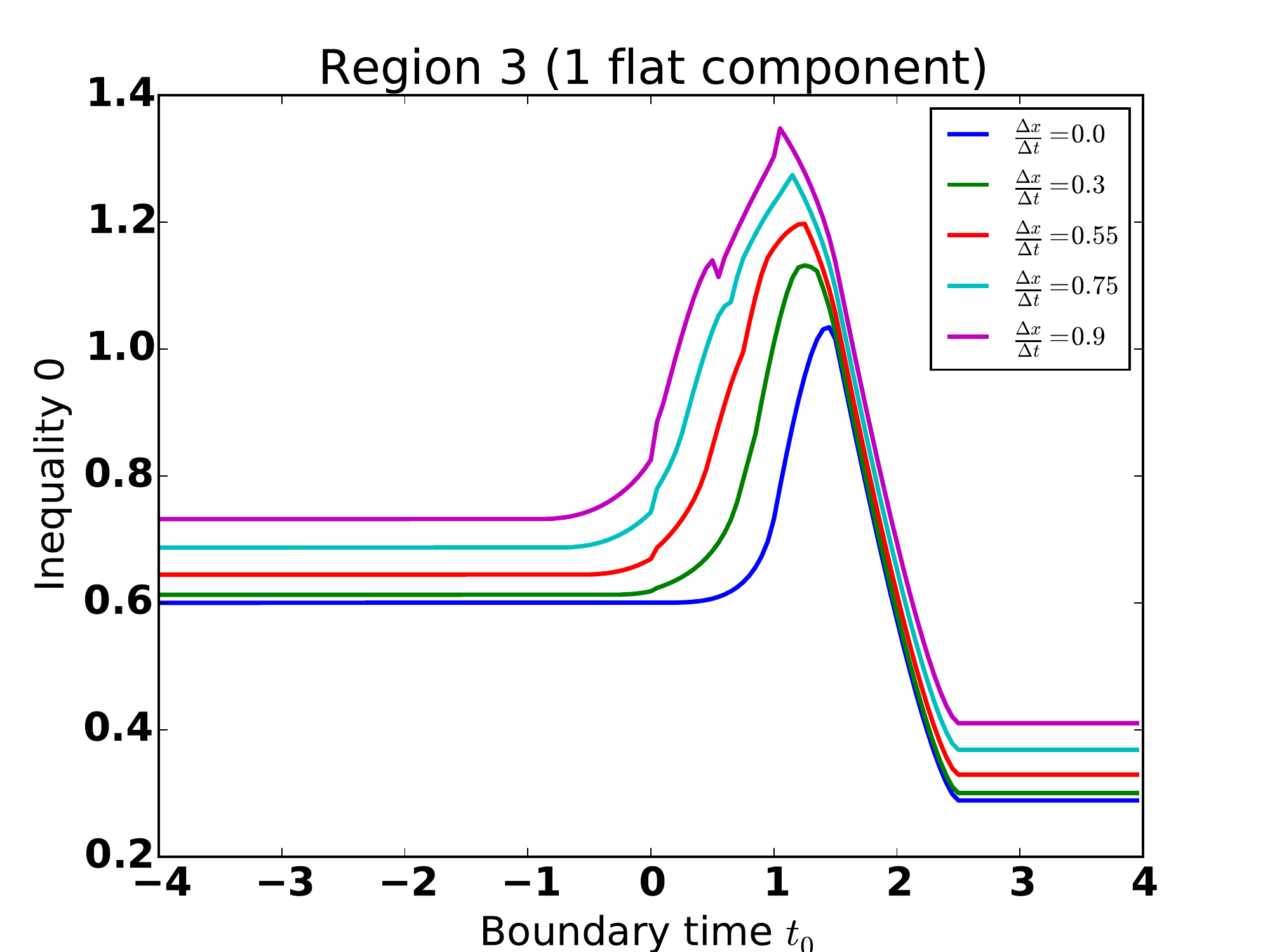}

 \caption{Inequality 0 plotted vs $t_b$. }
 \end{subfigure}

\begin{subfigure}{0.45\textwidth}
 \includegraphics[width=\textwidth]{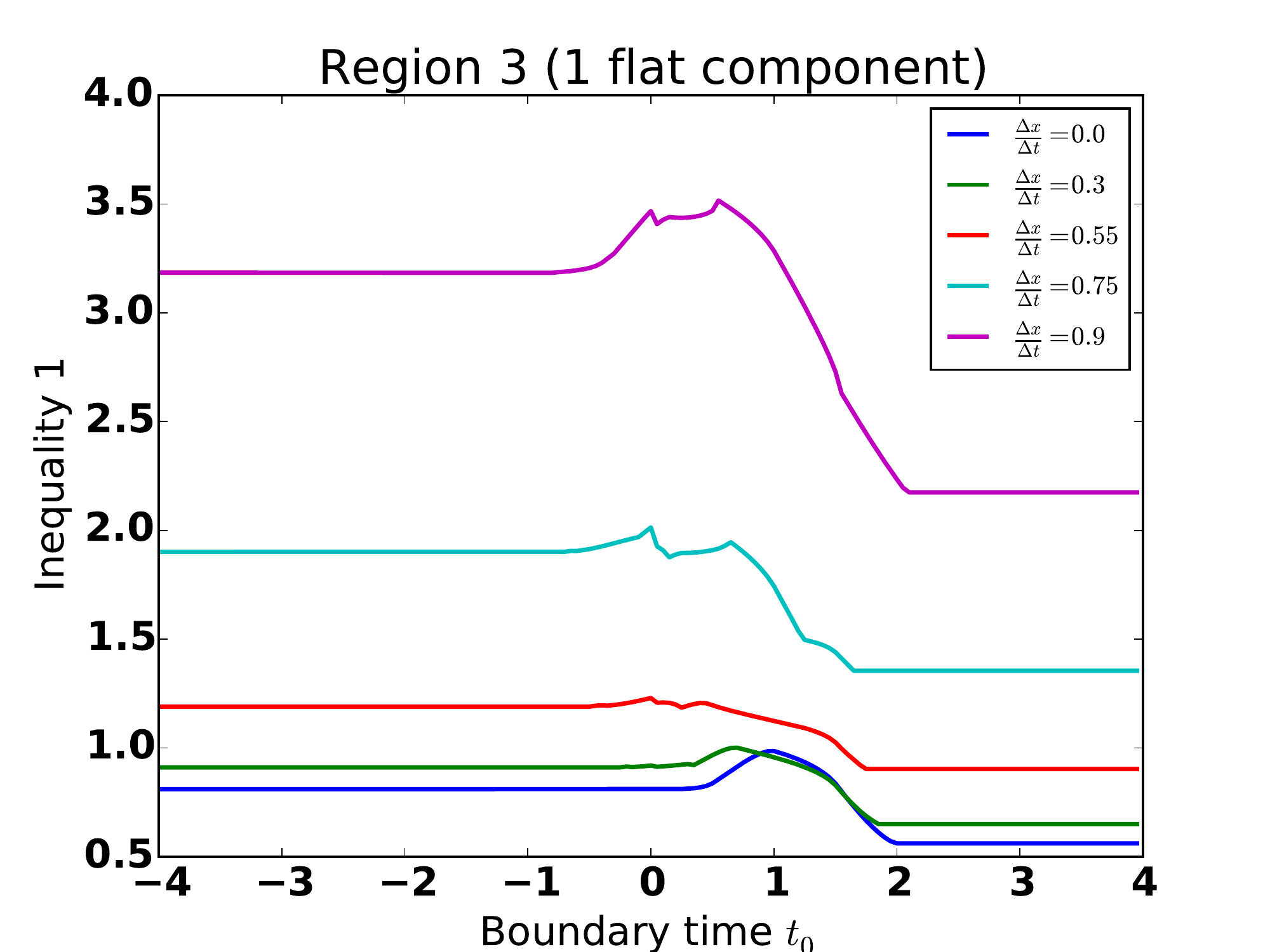}

 \caption{Inequality 1 plotted vs $t_b$. }
 \end{subfigure}
 \begin{subfigure}{0.45\textwidth}
 \includegraphics[width=\textwidth]{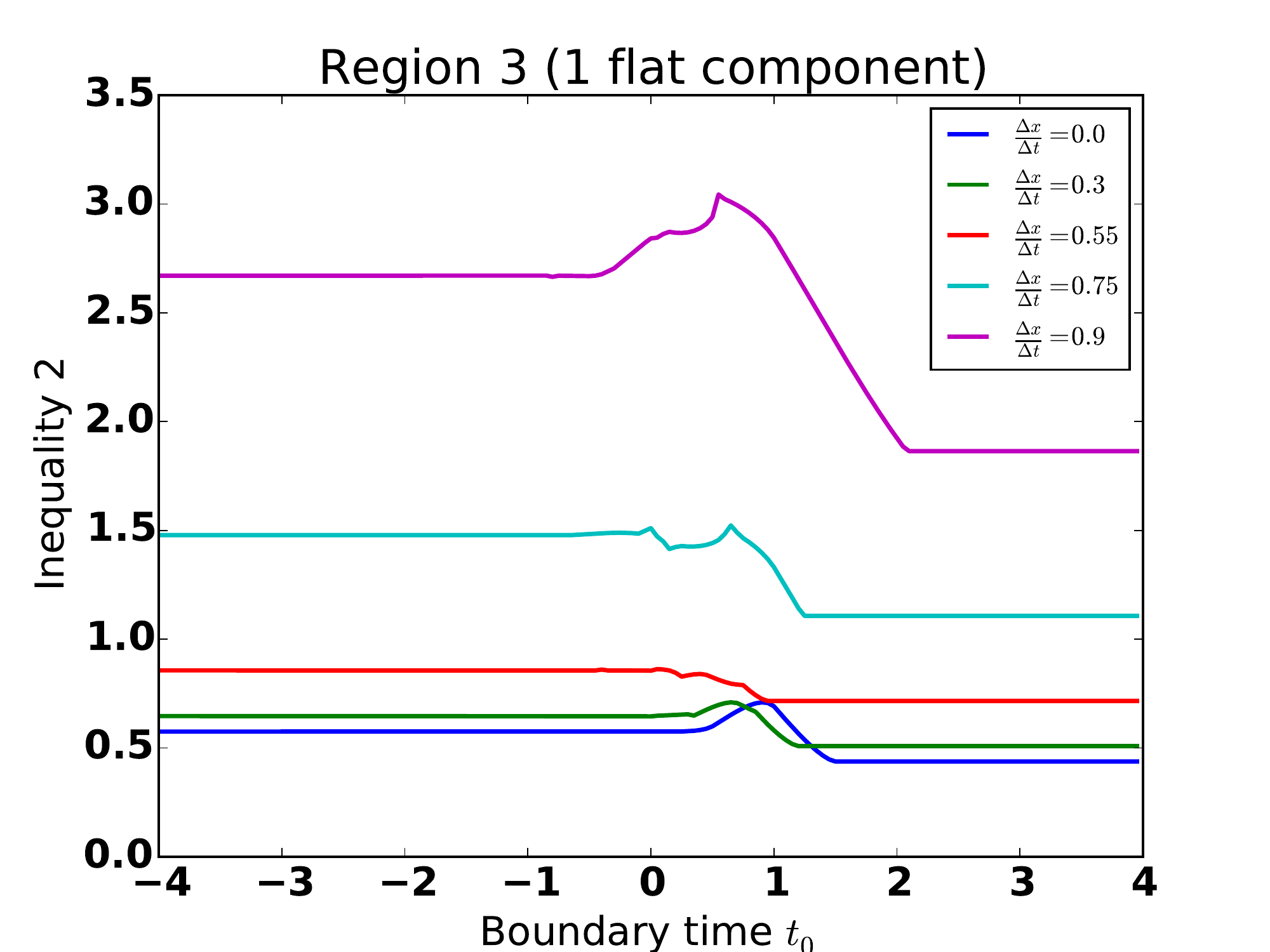}

 \caption{Inequality 2 plotted vs $t_b$. }
 \end{subfigure}
 \begin{subfigure}{0.45\textwidth}
 \includegraphics[width=\textwidth]{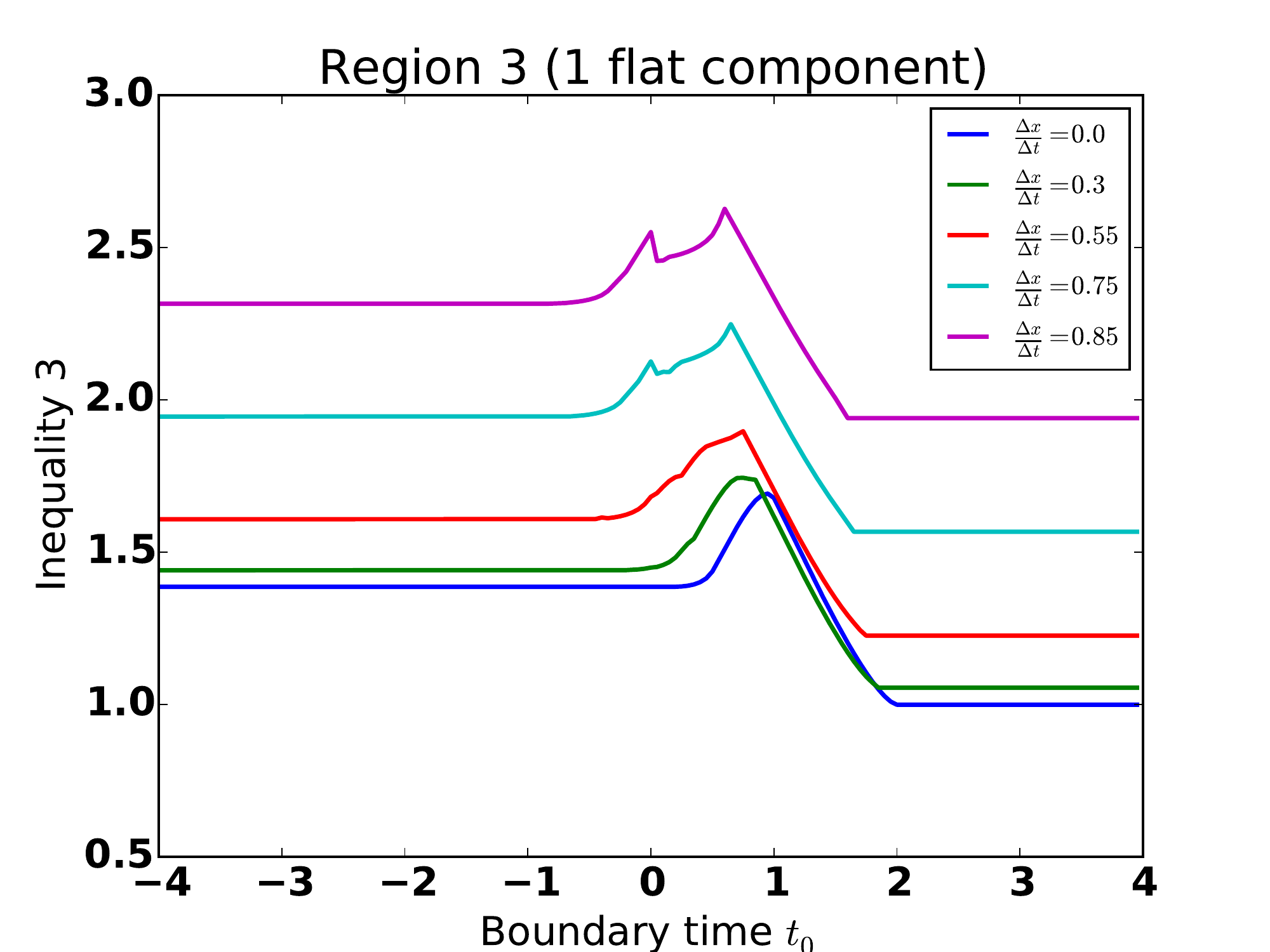}

 \caption{Inequality 3 plotted vs $t_b$. }
 \end{subfigure}
 \begin{subfigure}{0.45\textwidth}
 \includegraphics[width=\textwidth]{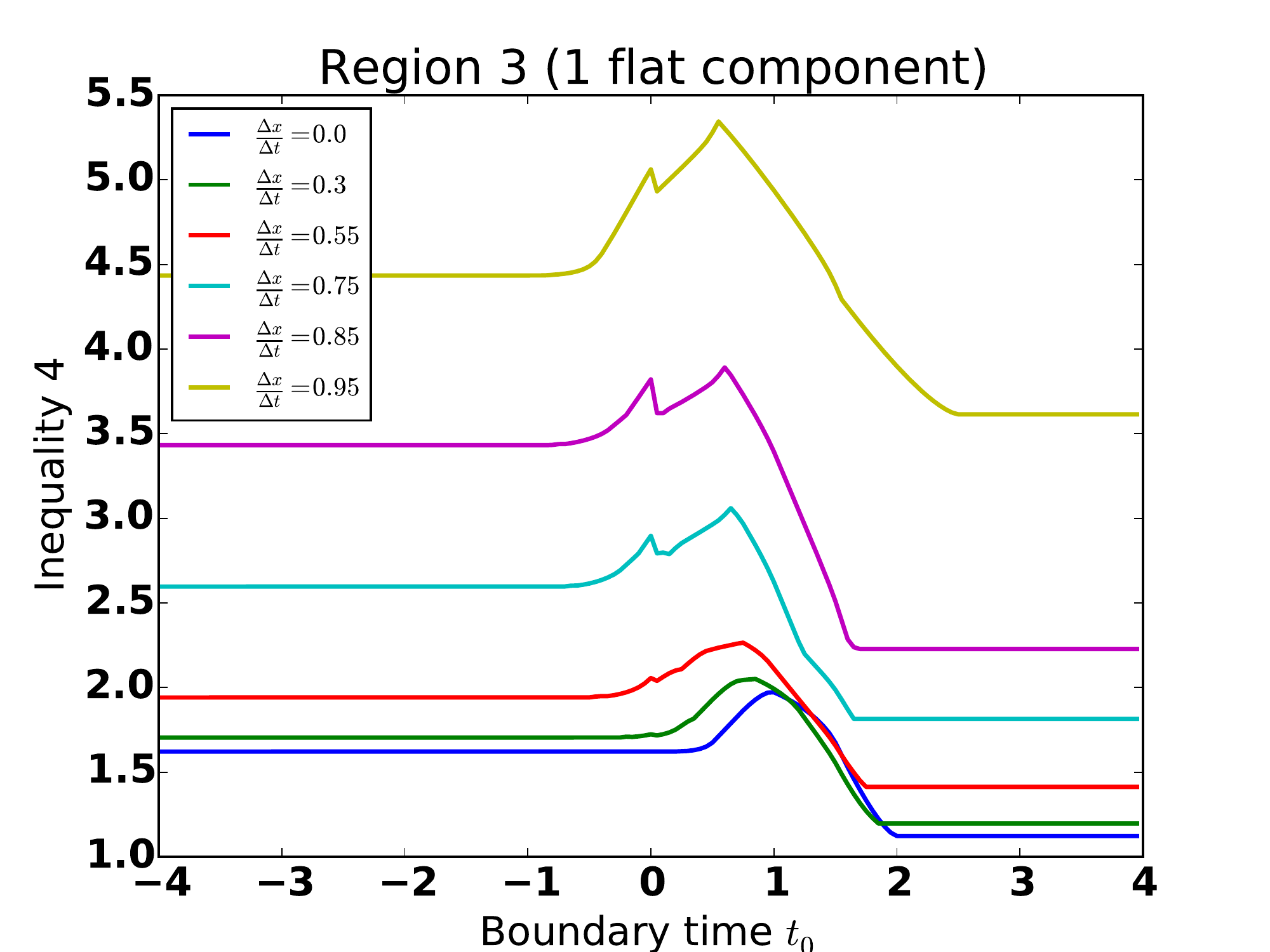}

 \caption{Inequality 4 plotted vs $t_b$. }
 \end{subfigure}
 
 \caption{The inequalities plotted vs $t_b$ for the Region 3 (with 1 flat component) for a variety of values of $\Delta t / \Delta x$. We see that the inequalities are all satisfied.}\label{fig:five_flat1}
\end{figure}

We now test the five-region inequalities. We use the same labeling scheme for the inequalities as used above, in the constant-time case. We consider the three configurations shown in Figure~\ref{fig:regions}. We consider a variety of values of $\Delta t/ \Delta x$, again fixing the value of $\Delta x$ for each component to be 1.  We plot these curves as functions of the boundary start time $t_b$. We do this for the zigzag configuration in Figure~\ref{fig:five_Z}, the configuration with 2 flat components in Figure~\ref{fig:five_flat2}, and the configuration with 1 flat component in Figure~\ref{fig:five_flat1} We see that the inequalities are all satisfied, and that the shapes of the curves strongly resemble those of the strong subadditivity. 

\section{Negative Energy Vaidya Metric}

We now consider the negative-energy Vaidya metric. As discussed above, this violates the null energy condition. We will see that strong subadditivity is violated, as well as the five-body inequalities. 

\subsection{Geodesic Kinematics} 

We consider constant-time intervals. For boundary time $t_b<0$ the geodesics will be entirely in the BTZ bulk, while for large enough $t_b$ it will be entirely in the AdS bulk. These cases were treated above; we will now consider the case where the geodesic is partially in the AdS region, and partially in the BTZ region. We first consider the BTZ part. By the symmetry of the problem, $E=0$ in the BTZ arc. Suppose that the geodesic crosses the shell at $r_c$. The value of the affine parameter at this value of $r$ is 
$$\tau_c = \log (\sqrt{ \abs{r_c^2-1}}+\sqrt{ \abs{r_c^2-p_x^2}}).$$
From the equation for $\dot{r}$, it is clear that $r=p_x$ is the turning point. Therefore, by symmetry, the length of the BTZ part of the geodesic is 
$$ L_B = 2 ( \tau_c -\tau_{p_x} ) = 2 \log \left ( \frac{\sqrt{ \abs{r_c^2-1}}+\sqrt{ \abs{r_c^2-p_x^2}}}{\sqrt{ \abs{p_x^2-1}} }\right ).$$
Similarly, the change in $x$ is given by 
$$\Delta x_B= -\log \left ( \frac{(p_x-1)^2+(\sqrt{ \abs{r_c^2-1}}+\sqrt{ \abs{r_c^2-p_x^2}})^2}{(p_x+1)^2+(\sqrt{ \abs{r_c^2-1}}+\sqrt{ \abs{r_c^2-p_x^2}})^2}\right )+\log \left ( \frac{(p_x-1)^2+(\sqrt{ \abs{p_x^2-1}})^2}{(p_x+1)^2+(\sqrt{ \abs{p_x^2-1}})^2}\right ).$$

We now turn to the AdS components. Similar to the positive-energy case, we require that at the shell $v=0$ we must have 
$$v' = 2 (r_A ' -r_B').$$
$E_B=0$ so 
$$v'=\frac{r_B'}{r_c^2-1},$$
which means that 
$$ r_A' = r_B'+ \frac{v'}{2}=r_B'+ \frac{r_B'}{2(r_c^2-1)}=\frac{(2 r_c^2-1)r_B'}{2(r_c^2-1)}. $$
Note that this means that $r_A'$ becomes negative when $r_c<\frac{1}{2}$. 
In AdS, we know that 
$$r_A'^2= \frac{r_c^6}{p_x^2}+\frac{r_c^4 E_A^2}{p_x^2}-r_c^4,$$
$$r_A'^2 p_x^2= r_c^6+r_c^4 E_A^2-r_c^4 p_x^2,$$
$$\frac{r_A'^2 p_x^2}{r_c^4}+p_x^2- r_c^2= E_A^2,$$
$$E_A^2=\frac{(2 r_c^2-1)^2 r_B'^2 p_x^2}{4 r_c^4 (r_c^2-1)^2}+p_x^2- r_c^2.$$
Also, we know that 
$$r_B'^2 = (r_c^2-1)r_c^2 (\frac{r_c^2}{p_x^2}-1 ),$$
which means 
$$E_A^2=\frac{(2 r_c^2-1)^2 (r_c^2- p_x^2)}{4 r_c^2 (r_c^2-1)}+p_x^2- r_c^2=\frac{((2 r_c^2-1)^2 -4r_c^4+4r_c^2)(r_c^2- p_x^2)}{4 r_c^2 (r_c^2-1)}=\frac{(r_c^2- p_x^2)}{4 r_c^2 (r_c^2-1)}.$$
If $r_c>\frac{1}{2}$ then $r'>0$ and $\dot{r}>0$, and the solution is 
$$ r(\tau ) = \frac{1}{2} (e^{\tau}+(p_x^2-E^2 ) e^{-\tau}).$$
We have that 
$$ \tau_c = \log (r_c + \sqrt{r_c^2+E_A^2 - p_x^2} ).$$
For large $R$, the corresponding (large, positive) value of $\tau$ is $\log 2R$ so the length of the AdS arc is 
$$L_A = \log 2R - \log (r_c + \sqrt{r_c^2+E_A^2 - p_x^2} ).$$
The change in $x$ is given by 
$$\Delta x_A = \frac{p_x}{r_c (r_c +\sqrt{r_c^2+E_A^2-p_x^2})}.$$
We find the starting point of the interval in the usual way:
$$t_b = t_c -\Delta t_A = \frac{1}{r_c} -\frac{E_A}{r_c (r_c +\sqrt{r_c^2+E_A^2-p_x^2})},$$
since $\Delta t_A = \Delta x_A \frac{E_A}{p_x}$. The total arc length and displacement are 
$$L = L_B +2L_A, \Delta x = 2 \Delta x_A+ \Delta x_B.$$

On the other hand, if $r_c< \frac{1}{2},$ $\dot{r}<0$ and the solution is given by
$$ r(\tau ) = \frac{1}{2} (e^{-\tau}+(p_x^2-E^2 ) e^{\tau}),$$
and we have that 
$$ \tau_c = - \log (r_c + \sqrt{r_c^2+E_A^2 - p_x^2} ).$$
The positive affine parameter for large $R$ is given by 
$$\tau_\infty = \log (2R) - \log (p_x^2 -E_A^2),$$
which means that the total length of the AdS arc is given by 
$$L_A = \tau_\infty-\tau_c = \log(2R) - \log (p_x^2 -E_A^2) + \log(r_c + \sqrt{r_c^2+E_A^2 - p_x^2} ).$$
Meanwhile, 
$$\Delta x_A = x(\tau_\infty) -x(\tau_c)= \frac{p_x}{r_c (r_c +\sqrt{r_c^2+E_A^2-p_x^2})}-\frac{2p_x}{-E_A^2+p_x^2+(p_x^2-E_A^2)^2}$$
and 
$$t_b=\frac{1}{r_c} - \frac{E_A}{p_x} \Delta x_A.$$
Once again, the total geodesic length and displacement are 
$$L = L_B +2L_A, \Delta x = 2 \Delta x_A+ \Delta x_B.$$

\begin{figure}
\centering
 \includegraphics[width=0.7\textwidth]{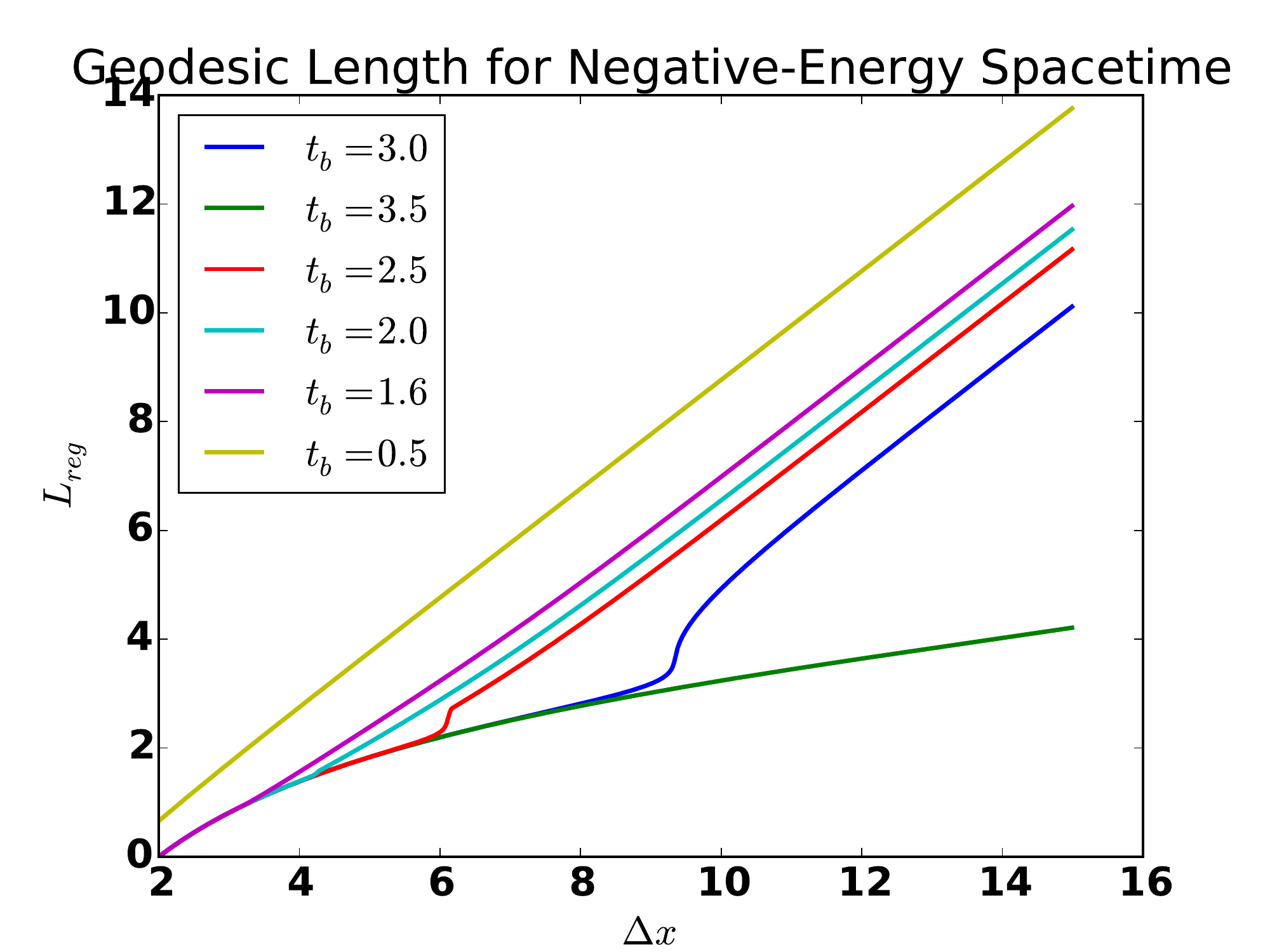}

 \caption{The regularized geodesic length $L_{reg}$ as a function of the boundary interval length, $\Delta x$ for various values of the boundary time $t_b$ in the negative-energy Vaidya spacetime. We see that the curves are not convex, meaning there will be violations of strong subadditivity}\label{fig:geo_len_NE}
\end{figure}

\subsection{Tests of Entropy Inequalities}

To find the geodesic for the negative-energy metric for a given interval, we proceed as follows. First, if $t_b \leq 0$, the geodesic is of course entirely in BTZ. If $t_b >0$, we find the trajectory of the geodesic in AdS, and calculate $v(\tau)$. If at any point it dips below 0, then there will be a portion of the geodesic that is in the BTZ bulk. We then use a numerical algorithm to find the values of $r_c$ and $p_x$ that correspond to the desired $t_b$ and $\Delta x$. We show a plot of the (regularized) geodesic length as a function of the displacement for various values of the boundary time $t_b$ in Figure~\ref{fig:geo_len_NE}. We see the non-convex behavior of some of these curves, which means that strong subadditivity will be violated. 
\begin{figure}
\centering
 \includegraphics[width=0.7\textwidth]{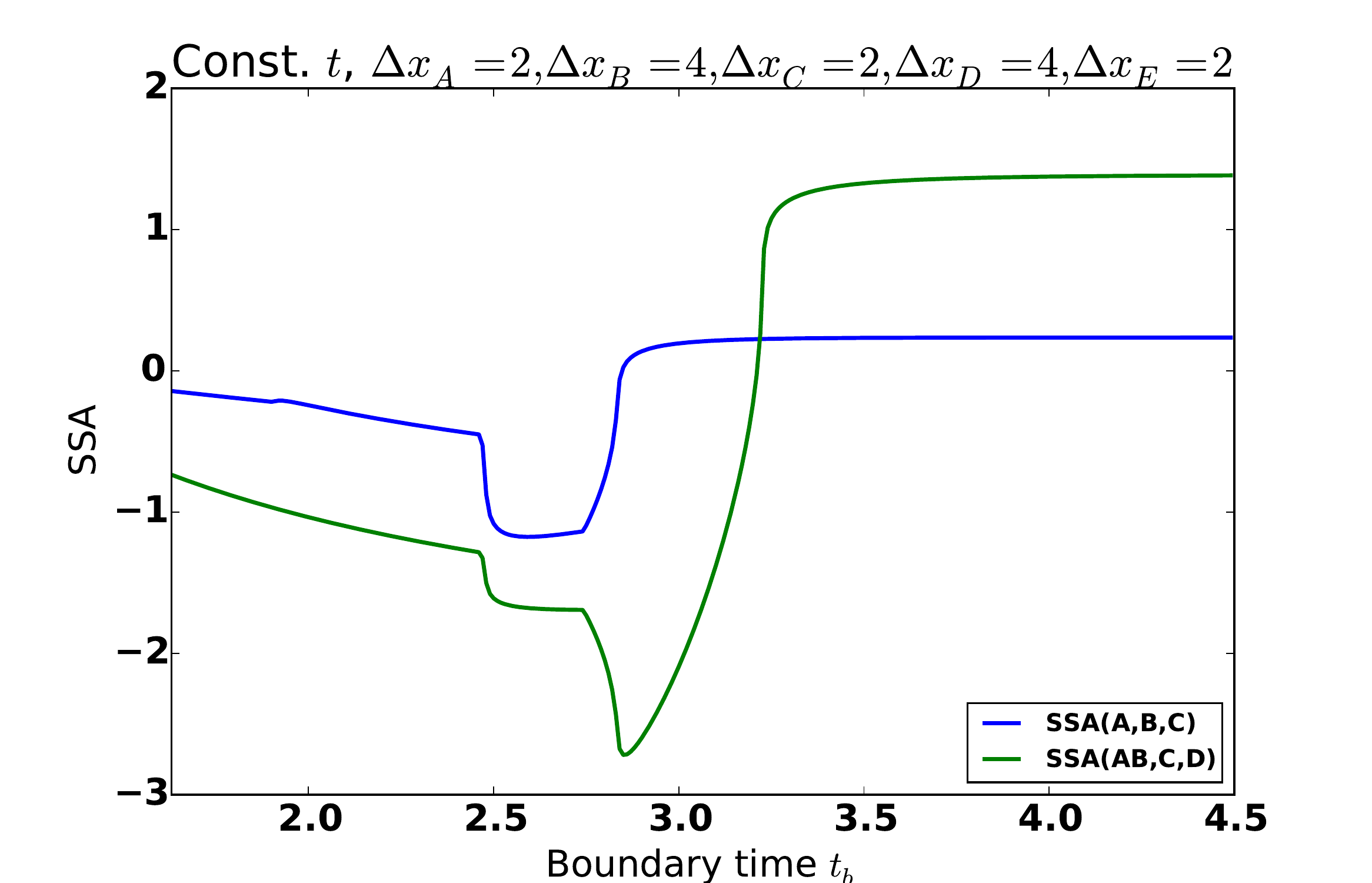}

 \caption{Strong subadditivity versus boundary time $t_b$ for the negative-energy Vaidya spacetime. We see that strong subadditivity is violated. Here, we consider strong subadditivity for regions $A, B,$ and $C,$ as well as for $AB, C,$ and $D.$ }\label{fig:SSA_NE}
\end{figure}

We consider five adjacent constant-time intervals, $A,B,C,D,E$. $A,C,$ and $E$ have width 2, while $B$ and $D$ have width 4. To start with, we plot strong subadditivity for a couple collections of regions in Figure~\ref{fig:SSA_NE}.  We see that strong subadditivity is violated, which is expected since our metric violates the null energy condition. 

\begin{figure}

\centering

\begin{subfigure}{0.45\textwidth}
 \includegraphics[width=\textwidth]{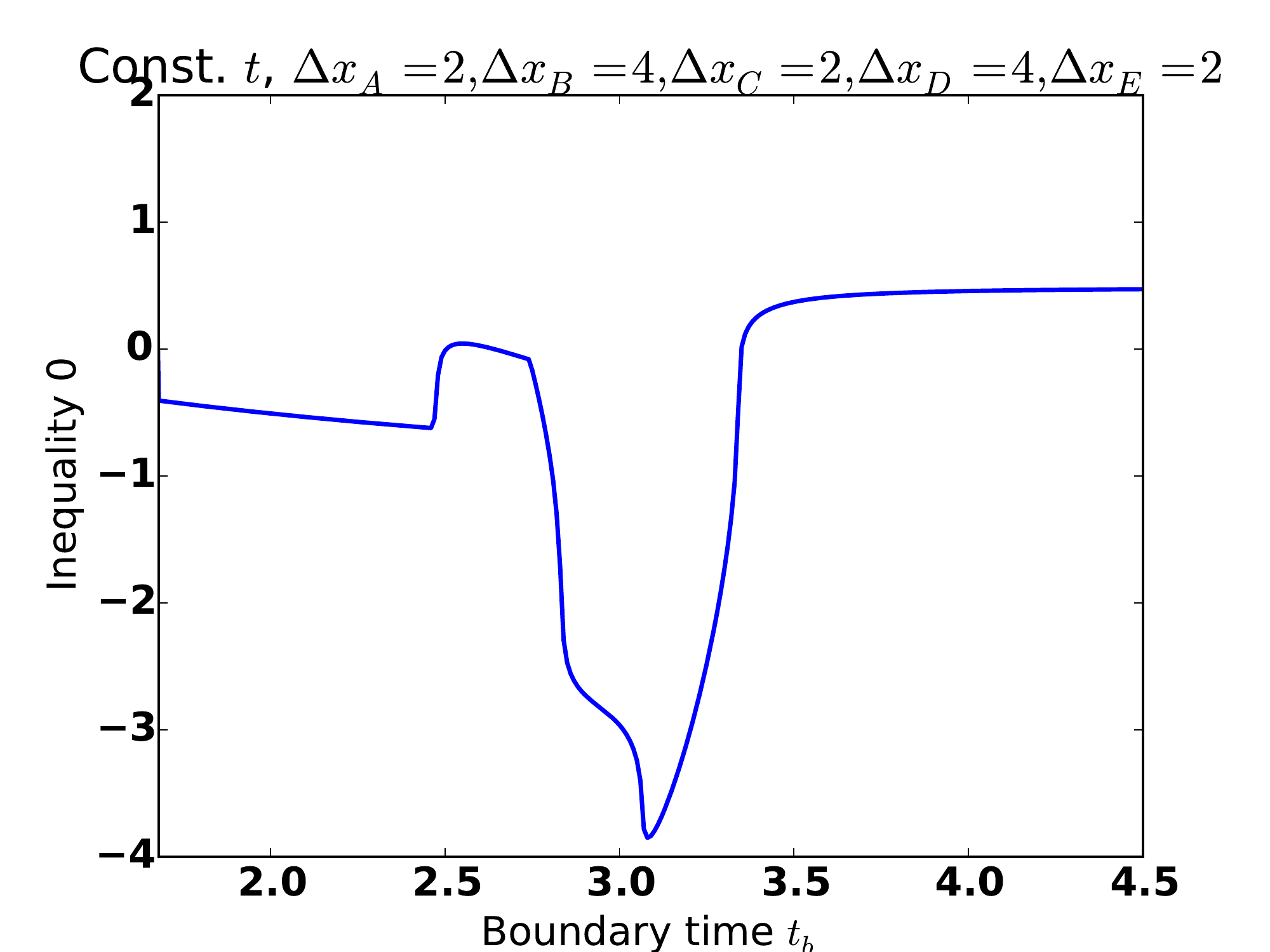}

 \caption{Inequality 0 plotted vs $t_b$. }
 \end{subfigure}

\begin{subfigure}{0.45\textwidth}
 \includegraphics[width=\textwidth]{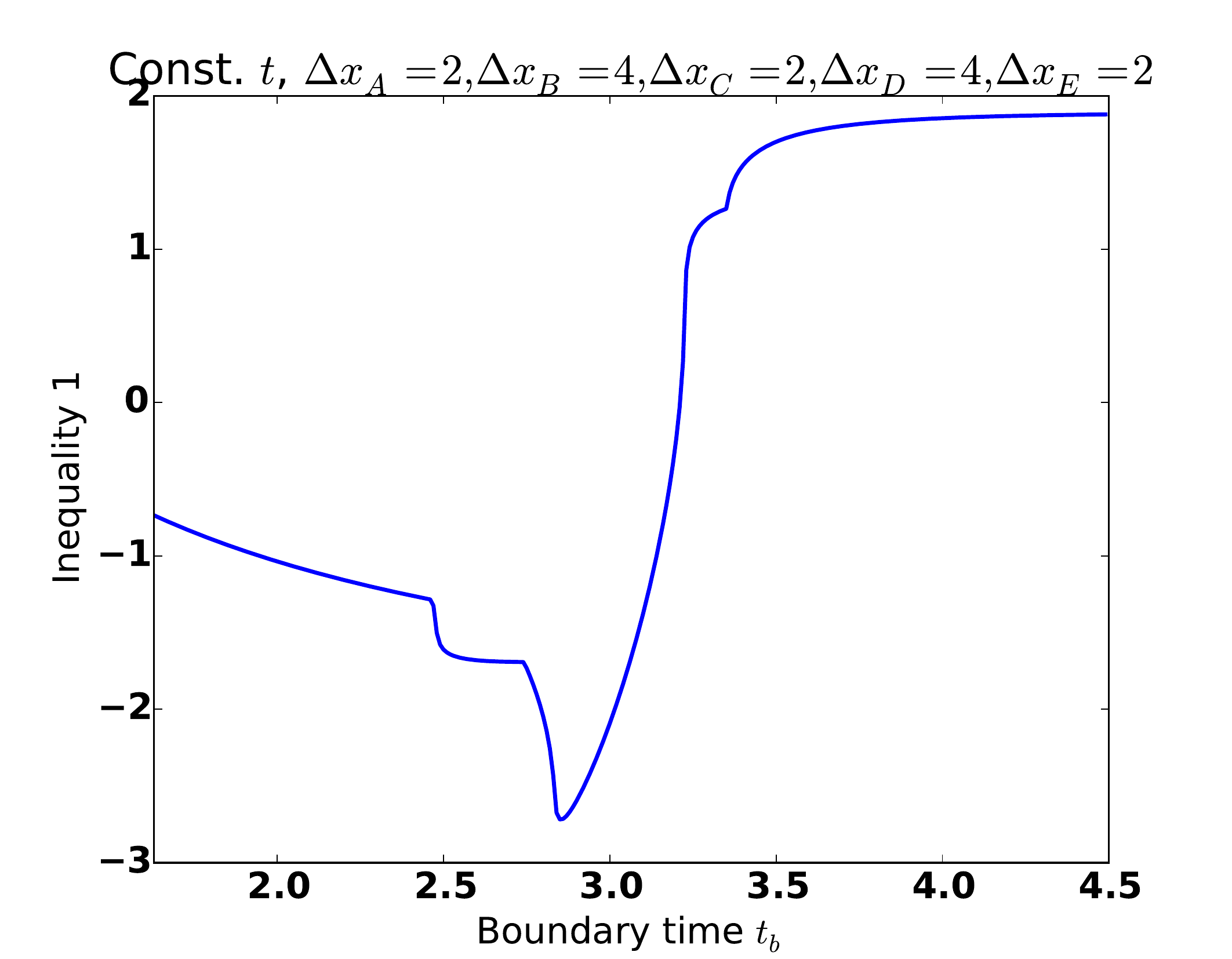}

 \caption{Inequality 1 plotted vs $t_b$. }
 \end{subfigure}
 \begin{subfigure}{0.45\textwidth}
 \includegraphics[width=\textwidth]{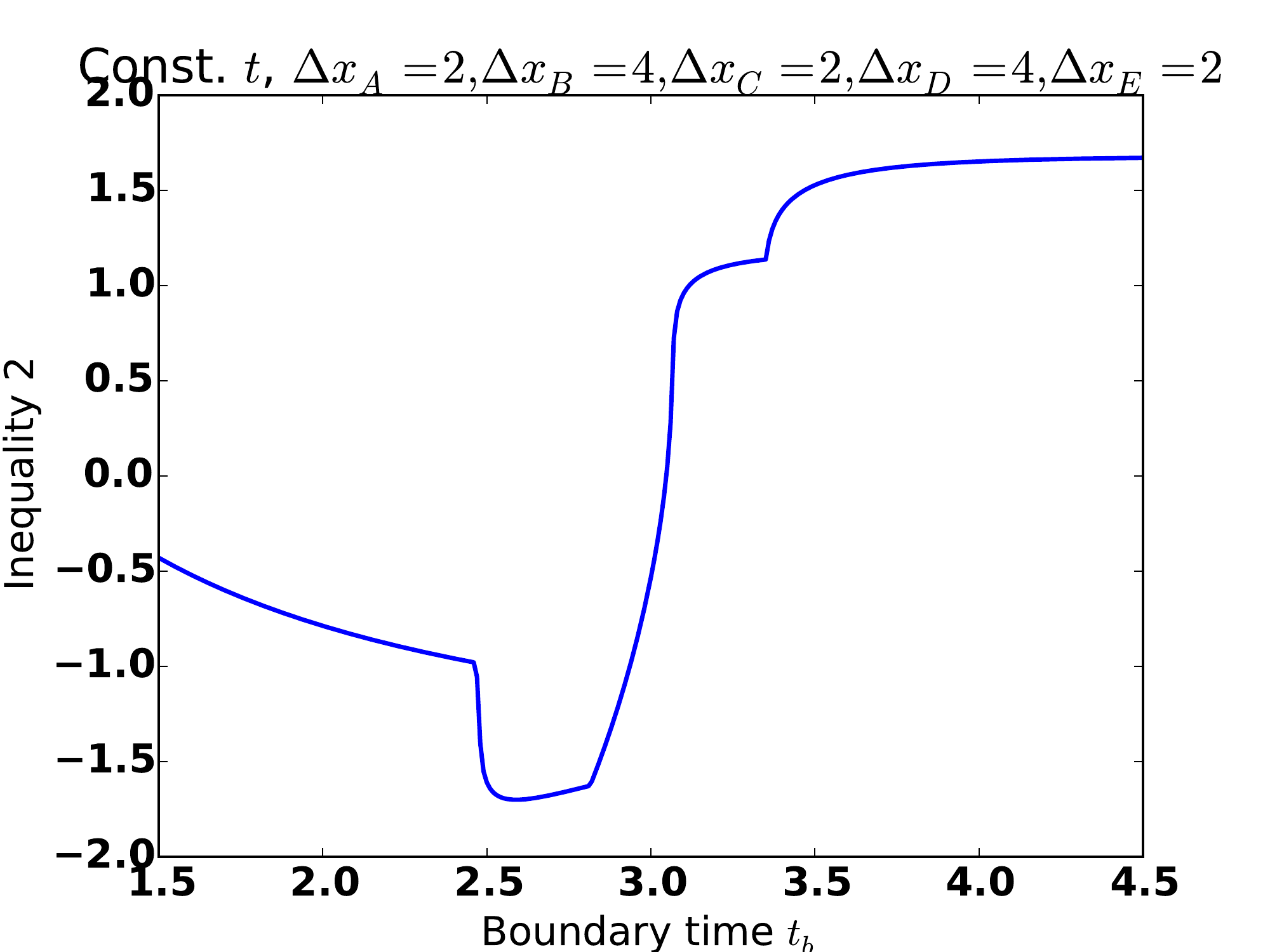}

 \caption{Inequality 2 plotted vs $t_b$. }
 \end{subfigure}
 \begin{subfigure}{0.45\textwidth}
 \includegraphics[width=\textwidth]{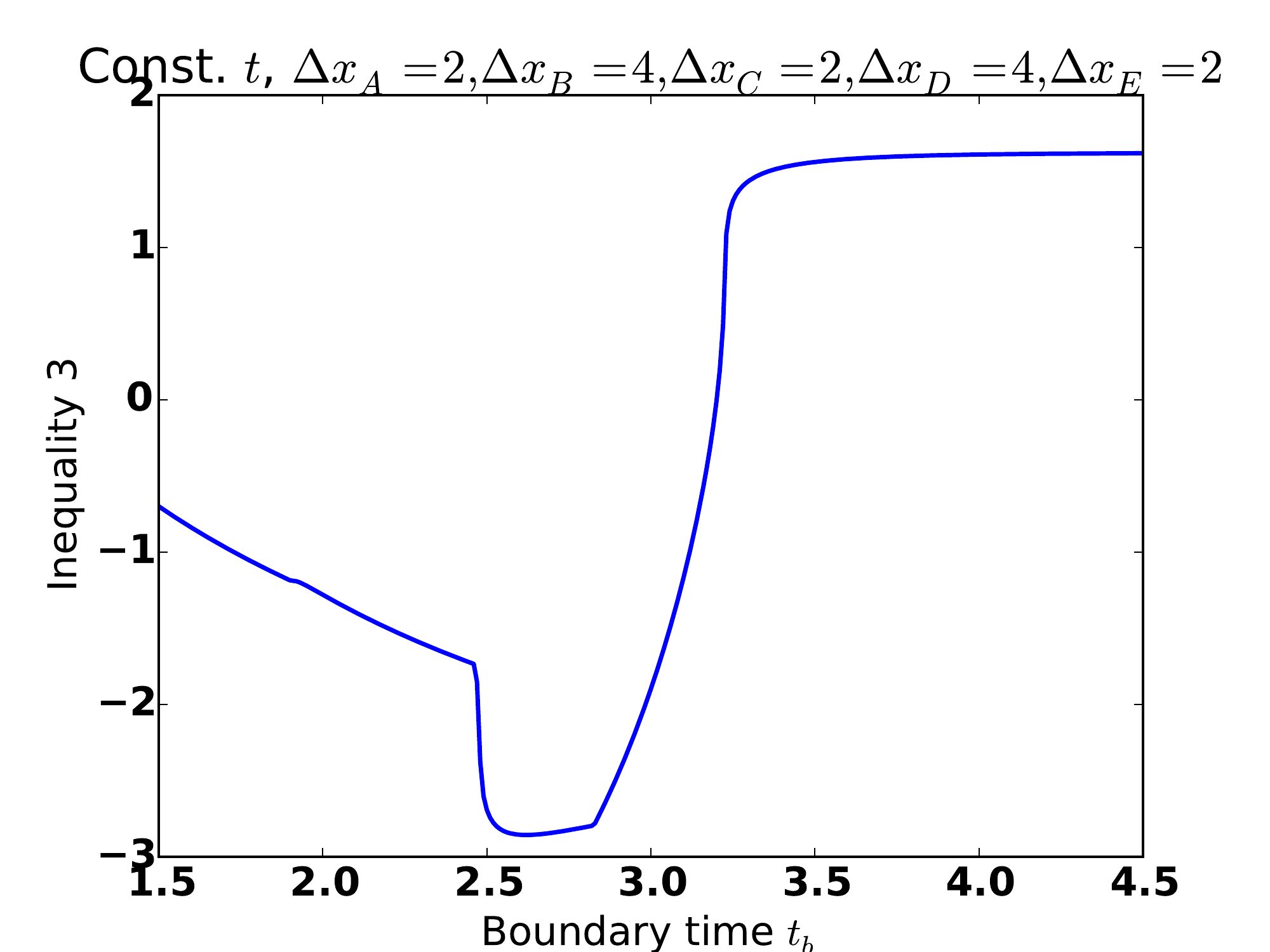}

 \caption{Inequality 3 plotted vs $t_b$. }
 \end{subfigure}
 \begin{subfigure}{0.45\textwidth}
 \includegraphics[width=\textwidth]{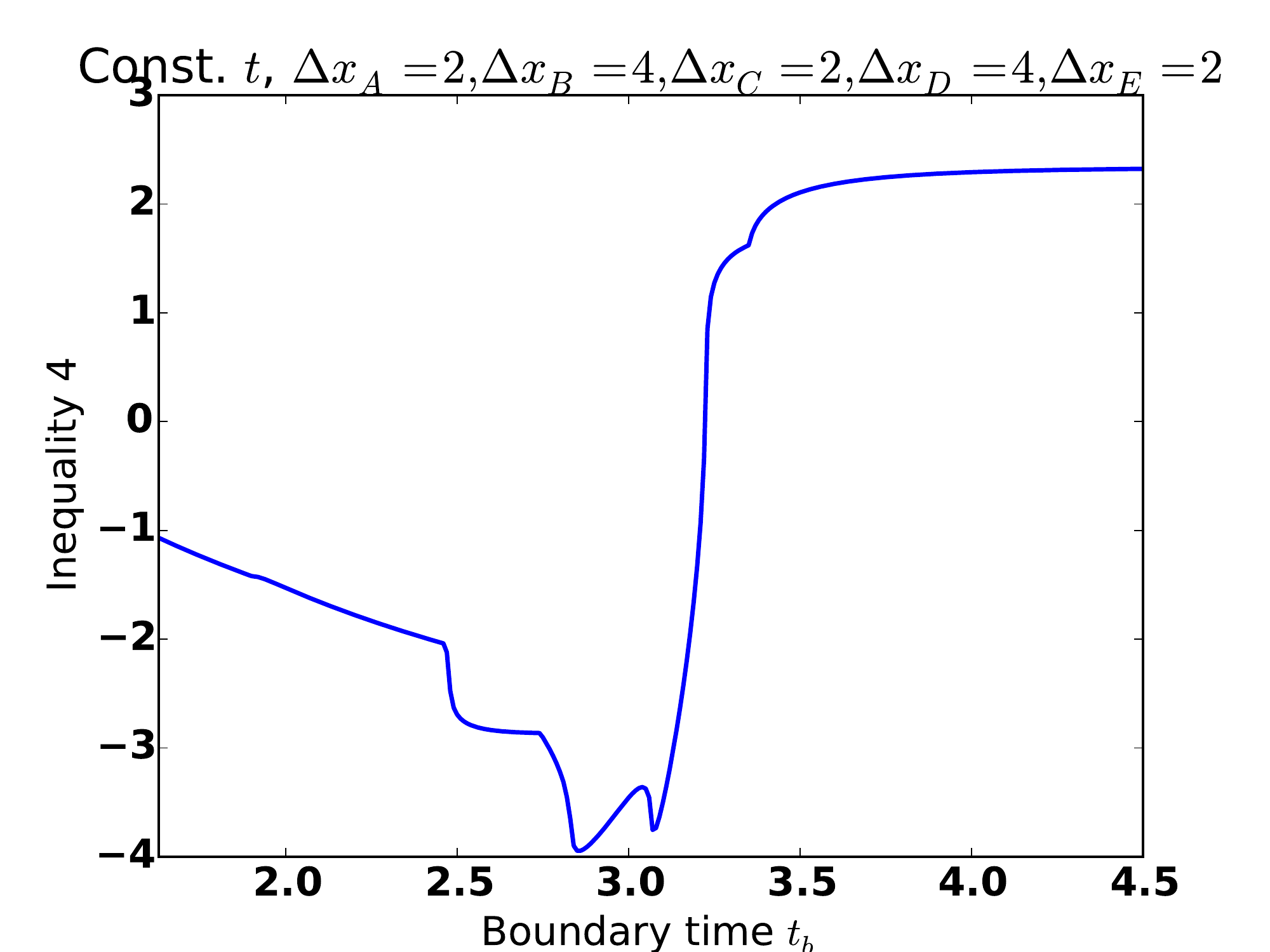}

 \caption{Inequality 4 plotted vs $t_b$. }
 \end{subfigure}
 
 \caption{The inequalities plotted vs $t_b$. We see that the inequalities are all violated for this spacetime that violates the null energy condition.}\label{fig:five_NE}
\end{figure}

Next, we check the five-region inequalities. We use the same numbering scheme as before (with labels 0 through 4), and we plot $4G_N$ times the left hand side minus $4G_N$ times the right hand side of each of the inequalities. We show the results in Figure~\ref{fig:five_NE}. We see that all of the inequalities are violated for this spacetime, roughly in the places where strong subadditivity is violated. Furthermore, we once again see that the curves for the five-region inequalities resemble the strong subadditivity curves.

%
%


\acknowledgments

I am very grateful to Matt Headrick for suggesting this problem and fruitful discussions, as well as useful comments on the manuscript. I would also like to thank Ning Bao, Alan Guth, Daniel Harlow, Mark Kon, Sam Leutheusser, and Hong Liu for useful conversations.



\begin{thebibliography}{99}

\bibitem{Maldacena} 
  J.~M.~Maldacena,
  ``The Large N limit of superconformal field theories and supergravity,''
  Int.\ J.\ Theor.\ Phys.\  {\bf 38}, 1113 (1999)
  doi:10.1023/A:1026654312961, 10.4310/ATMP.1998.v2.n2.a1
  [hep-th/9711200].
  
\bibitem{Gubser} 
  S.~S.~Gubser, I.~R.~Klebanov and A.~M.~Polyakov,
  ``Gauge theory correlators from noncritical string theory,''
  Phys.\ Lett.\ B {\bf 428}, 105 (1998)
  doi:10.1016/S0370-2693(98)00377-3
  [hep-th/9802109].
  
\bibitem{Witten} 
  E.~Witten,
  ``Anti-de Sitter space and holography,''
  Adv.\ Theor.\ Math.\ Phys.\  {\bf 2}, 253 (1998)
  doi:10.4310/ATMP.1998.v2.n2.a2
  [hep-th/9802150].
  
\bibitem{RT} 
  S.~Ryu and T.~Takayanagi,
  ``Holographic derivation of entanglement entropy from AdS/CFT,''
  Phys.\ Rev.\ Lett.\  {\bf 96}, 181602 (2006)
  doi:10.1103/PhysRevLett.96.181602
  [hep-th/0603001].
  
\bibitem{HRT} 
  V.~E.~Hubeny, M.~Rangamani and T.~Takayanagi,
  ``A Covariant holographic entanglement entropy proposal,''
  JHEP {\bf 0707}, 062 (2007)
  doi:10.1088/1126-6708/2007/07/062
  [arXiv:0705.0016 [hep-th]].

\bibitem{Headrick} 
  R.~Callan, J.~Y.~He and M.~Headrick,
  ``Strong subadditivity and the covariant holographic entanglement entropy formula,''
  JHEP {\bf 1206}, 081 (2012)
  doi:10.1007/JHEP06(2012)081
  [arXiv:1204.2309 [hep-th]].
  
\bibitem{RT_deriv} 
  A.~Lewkowycz and J.~Maldacena,
  ``Generalized gravitational entropy,''
  JHEP {\bf 1308}, 090 (2013)
  doi:10.1007/JHEP08(2013)090
  [arXiv:1304.4926 [hep-th]].
  
\bibitem{HRT_deriv} 
  X.~Dong, A.~Lewkowycz and M.~Rangamani,
  ``Deriving covariant holographic entanglement,''
  JHEP {\bf 1611}, 028 (2016)
  doi:10.1007/JHEP11(2016)028
  [arXiv:1607.07506 [hep-th]].
  
\bibitem{Lieb1} 
  E.~H.~Lieb and M.~B.~Ruskai,
  ``A Fundamental Property of Quantum-Mechanical Entropy,''
  Phys.\ Rev.\ Lett.\  {\bf 30}, 434 (1973).
  doi:10.1103/PhysRevLett.30.434
  
\bibitem{Lieb2} 
  E.~H.~Lieb and M.~B.~Ruskai,
  ``Proof of the strong subadditivity of quantum-mechanical entropy,''
  J.\ Math.\ Phys.\  {\bf 14}, 1938 (1973).
  doi:10.1063/1.1666274
  
\bibitem{RT_SSA} 
  M.~Headrick and T.~Takayanagi,
  ``A Holographic proof of the strong subadditivity of entanglement entropy,''
  Phys.\ Rev.\ D {\bf 76}, 106013 (2007)
  doi:10.1103/PhysRevD.76.106013
  [arXiv:0704.3719 [hep-th]].
  

  
\bibitem{RT_MMI} 
  P.~Hayden, M.~Headrick and A.~Maloney,
  ``Holographic Mutual Information is Monogamous,''
  Phys.\ Rev.\ D {\bf 87}, no. 4, 046003 (2013)
  doi:10.1103/PhysRevD.87.046003
  [arXiv:1107.2940 [hep-th]].
  
 \bibitem{MMI_SSA2} 
  M.~Headrick,
  ``General properties of holographic entanglement entropy,''
  JHEP {\bf 1403}, 085 (2014)
  doi:10.1007/JHEP03(2014)085
  [arXiv:1312.6717 [hep-th]].
  
\bibitem{Wall} 
  A.~C.~Wall,
  ``Maximin Surfaces, and the Strong Subadditivity of the Covariant Holographic Entanglement Entropy,''
  Class.\ Quant.\ Grav.\  {\bf 31}, no. 22, 225007 (2014)
  doi:10.1088/0264-9381/31/22/225007
  [arXiv:1211.3494 [hep-th]].

\bibitem{Cone} 
  N.~Bao, S.~Nezami, H.~Ooguri, B.~Stoica, J.~Sully and M.~Walter,
  ``The Holographic Entropy Cone,''
  JHEP {\bf 1509}, 130 (2015)
  doi:10.1007/JHEP09(2015)130
  [arXiv:1505.07839 [hep-th]].
  
  \bibitem{BaoMezei} 
  N.~Bao and M.~Mezei,
  ``On the Entropy Cone for Large Regions at Late Times,''
  arXiv:1811.00019 [hep-th].
  
 \bibitem{Rota} 
  M.~Rota and S.~J.~Weinberg,
  ``New constraints for holographic entropy from maximin: A no-go theorem,''
  Phys.\ Rev.\ D {\bf 97}, no. 8, 086013 (2018)
  doi:10.1103/PhysRevD.97.086013
  [arXiv:1712.10004 [hep-th]].
  
  \bibitem{Flory} 
  M.~Flory, J.~Erdmenger, D.~Fernandez, E.~Megias, A.~K.~Straub and P.~Witkowski,
 ``Time dependence of entanglement for steady state formation in AdS$_3$/CFT$_2$,''
  J.\ Phys.\ Conf.\ Ser.\  {\bf 942}, no. 1, 012010 (2017)
  doi:10.1088/1742-6596/942/1/012010
  [arXiv:1709.08614 [hep-th]].
  
 \bibitem{Cuenca} 
  S.~Hernandez Cuenca,
  ``The Holographic Entropy Cone for Five Regions,''
  arXiv:1903.09148 [hep-th].
  
  \bibitem{Allais} 
  A.~Allais and E.~Tonni,
  ``Holographic evolution of the mutual information,''
  JHEP {\bf 1201}, 102 (2012)
  doi:10.1007/JHEP01(2012)102
  [arXiv:1110.1607 [hep-th]].






\end{thebibliography}
\end{document}